\documentclass[showpacs,floatfix]{revtex4}
\usepackage[dvips]{graphicx}
\usepackage{epsfig}

\usepackage[cp1251]{inputenc}
\usepackage[english]{babel}
\usepackage{color}
\usepackage{bm}

\usepackage[T1]{fontenc}
\usepackage{xcolor}
\usepackage{lmodern}
\usepackage{listings}
\begin{document}

\title{Landau theory for smectic-A -- hexatic-B coexistence in smectic films}

\author{E. I. Kats$^{1}$, V. V. Lebedev$^{1}$, E. S. Pikina$^{1}$}

\affiliation{$^{1}$Landau Institute for Theoretical Physics, RAS, \\
142432, Chernogolovka, Moscow region, Russia}

\begin{abstract}

We explain theoretically peculiarities of the smectic $A$ -- hexatic $B$ equilibrium phase coexistence in a finite
temperature range recently observed experimentally in free standing smectic films [I.A.Zaluzhnyy et al., Physical Review E,
{\bf 98}, 052703 (2018)]. We quantitatively describe this unexpected phenomenon within Landau phase transitions theory
assuming that the film state is close to a tricritical point. We found that the surface hexatic order leads to diminishing the
phase coexistence range as the film thickness decreases shrinking it at some minimal film thickness $L_c$, of the order
of the hexatic correlation length. We established universal laws for the temperature width of the phase coexistence in terms of
the reduced variables. Our theory is in agreement with the existing experimental data.


\end{abstract}

\maketitle

%\pacs{64.70.Md, 61.30.Dk, 61.30.Gd}
%\date{}

\maketitle

\section{Introduction}
\label{sec:int}

Free standing smectic films are unique layered systems, solid-like in one direction (normal to the layers) and fluid-like in two lateral directions.
Unlike other films, smectic films are living in three-dimensional world without any parasitic influence from a substrate. It is not surprising that
this topic is the subject of many experimental and theoretical works (see, e.g., the comprehensive review \cite{JO03} and the monograph
\cite{PP06}). Our motivation to add one more article to the investigation field is related to new results, concerning the phase coexistence
in smectic films, we obtained. Our main concern is related to finite-size effects.

In our study we developed the quantitative theory explaining the finite temperature interval for the equilibrium coexistence of
the smectic $A$ and the hexatic $B$ phases in the smectic films. Common wisdom claims that the equilibrium phase coexistence at the
first order phase transition takes place at the transition temperature solely. A finite range for the phase coexistence can be achieved for
binary mixtures, or (in the case of single-component materials) in confined geometry, where neither of the coexistent states can provide
the required equilibrium density. However, we consider one-component material and in apparently unconfined free standing film geometry.
The point is that any smectic liquid crystal is strongly anisotropic and solid-like along the normal to smectic layers. Due to this anisotropy,
smectic stress tensor component orthogonal to the smectic layers, is not determined uniquely by the external pressure,
an essential contribution comes from the solid-like elasticity of smectic layers (see more details in Refs. \cite{JO03,PP06}).
As a result, smectic films  behave similarly to a closed volume system undergoing the first order phase transition
under condition that the number of the smectic layers is fixed (i.e., unchanged on a time scale needed to get the equilibrium phase coexistence).
The standard experimental technique for the free standing smectic film preparation, indeed, provides the uniform film thickness
\cite{Stoebe94,Dolganov95,Picano01,Ostrovskii03,ZK18}. In turn, local changes in the film thickness are possible only under overheating of
the free standing smectic film above the bulk temperature of melting of smectic  phase or under local (nonuniform) heating of the films,
see Refs. \cite{Stoebe94,Dolganov95,Picano01,Stoebe95,Demikhov95,Huang97,Ostrovskii04,Stannarius08,PO15,PO17}. These non-equilibrium
phenomena are beyond our consideration.

In the majority of the materials, exhibiting $Sm A$ -- $Hex B$ phase transition, the transition turns out to be weak first order phase transitions, see Refs. \cite{JV95,HK97,RD05,MP13,ZK18}. A tempting explanation of the fact based on closeness to a critical point, is excluded since the states  have different symmetries ($Sm A$ possesses isotropic liquid-like smectic layers, and $Hex B$ possesses orientation hexagonal symmetry order). We suggest another possibility to explain the experimental data, that the system is close to a tricritical point. This assumption is supported also by the
measured critical exponents (for the specific heat and for the order parameter) that are close to those for the tricritical point \cite{JV95,HK97,RD05,MP13}. It is worth to note that the liquid-crystalline materials exhibiting the $Sm A$ -- $Hex B$ phase transition demonstrate apparently universal behavior. The phase diagrams of such materials are remarkably similar even though the molecules of the
materials are appreciably different (see, for example, \cite{JV96,ZK17}). Thus our results are universal and can be applied to all the materials.

We exploit the phenomenological Landau phase transitions theory. As it is known, the mean field Landau theory works well near the tricritical
point (up to logarithmic corrections), see, e.g., \cite{LL80,ST87,AN91}. Our calculations are mainly analytical, giving the frame for observable effects. They are expressed as universal laws in terms of reduced variables. To find solutions of the non-linear equations within the whole temperature interval of the phase coexistence we use Wolfram Mathematica numerics. This allow us to illustrate dependencies for the width of the equilibrium phase coexistence on system parameters. We also compute numeric values of the dimensionless coefficients entering the
derived analytically universal laws.

In the work \cite{ZK18} the coexistence of $Sm A$ and $Hex B$  phases were observed  in a finite temperature interval (and qualitatively
and semi-quantitatively, for thick films, rationalized theoretically). However, presented in the paper \cite{ZK18} the expression for the temperature interval of the equilibrium phase coexistence  has been derived merely from the surface order induced renormalization of the bulk hexatic phase parameters (what is not a consistent procedure). In this work we present the consistent quantitative theory.

Our paper is organized as follows. In the next section \ref{sec:bulk} we formulate general thermodynamical conditions for the phase
coexistence, in a form suitable for smectic liquid crystals, possessing the layer structure. In the subsection \ref{sec:landau}
of this section \ref{sec:bulk}  we discuss the phase coexistence in bulk in terms of the Landau theory. Specifically motivated by experimental observations \cite{ZK18} we study $Sm A$ -- $Hex B$ transition in the free standing films. In the section \ref{sec:surface} we explore and analyze the key point of our work, namely the surface effects. In the free standing smectic films exhibiting $Sm A$ -- $Hex B$ phase transition, the surface hexatic order occurs at the temperature higher than the bulk transition temperature. This surface induced order in the vicinity of a tricritical point
penetrates into the interior of the film, what essentially influences the phase transition even for the relatively thick films. In particular we
demonstrate that the surface order provokes diminishing the phase coexistence range as the film thickness decreases. Eventually, it leads to
shrinking the coexistence range at some minimal film thickness $L_c$. Thus we arrive at a special critical point, where the coexisting phases
become indistinguishable. In the concluding section \ref{sec:con}, we summarize our results, and also discuss some open questions and
perspectives. We relegate some technical details of the analytic calculations into two appendices to the main text.

\section{General thermodynamic analysis of the phase coexistence}
\label{sec:bulk}

Here, we remind the general thermodynamical conditions of the phase coexistence \cite{LL80,HU87}. Two phase coexistence  indicates that
none of the coexisting phases (in our case $Sm A$ and $Hex B$) is able to support the optimal two-dimensional density of the film,
and the compromise is achieved by means of two-phase equilibrium where two phases coexist. The coexistence signals about first order
transition between the phases. However, ordering in the hexatic case is weak in the region of the phase coexistence.
That enables one to use the Landau expansion in the order parameter to analyze the phenomenon. We consider the case where
the number of the smectic layers in the film is fixed. The assumption holds if nucleation of dislocation loops
(which are able to adjust the number of layers to the external stresses) is very infrequent and too
slow in comparison to characteristic time scales relevant for the phase coexistence \cite{PP01,PP06,OP18}.
Then the thickness $L$ of the film is determined as a minimum condition of an appropriate thermodynamic potential.
Although the thicknesses of the $Sm A$ and $Hex B$ phases are slightly different (at a given temperature),
due to the weakness of the hexatic ordering the difference is small and can be safely neglected.

We designate as $N_H$ and $N_A$ two-dimensional mass densities and designate as $F_H$ and $F_A$
two-dimensional free energy densities of the $Hex B$ and of the $Sm A$ phases, respectively. If areas that the phases
occupy are $A_H$ and $A_A$, then the total free energy of the system can be written as
 \begin{eqnarray}
 {\cal F}= F_A(N_A) A_A+ F_H(N_H) A_H
 \nonumber \\
 -\mu (N_A   A_A +N_H A_H-{\cal N}),
 \label{ther1}
 \end{eqnarray}
where ${\cal N}$ is the total number of molecules in the film and $\mu$ is a Lagrangian multiplier fixing the number.
Minimization of the energy (\ref{ther1}) in terms of $N_A$ and $N_H$ leads to the conditions
 \begin{equation}
 \frac{\partial F_A}{\partial N_A}=\mu =
 \frac{\partial F_H}{\partial N_H}.
 \label{ther2}
 \end{equation}
Thus the chemical potentials of the phases are equal if they coexist. This condition is analogous to famous Maxwell common
tangent construction, see Refs. \cite{HU87,ST87,AN91,CL00}.

Note that $A_H=A-A_A$, where $A$ is the total area of the film. Therefore minimization of the expression (\ref{ther1}) in terms of $A_A$
leads to the condition
 \begin{equation}
 \Omega_A=\Omega_H,
 \label{ther3}
 \end{equation}
where $\Omega$ is the grand thermodynamic potential per unit area:
 \begin{eqnarray}
 \Omega(\mu)=F-N\frac{\partial F}{\partial N}, \quad
 \mu =\frac{\partial F}{\partial N},
 \nonumber \\
 d\Omega=-N d\mu -SdT.
 \label{grand}
 \end{eqnarray}
Further we operate in terms of the grand thermodynamic potential having in mind that both, the chemical potentials and the temperatures
of the coexisting phases should coincide.

We arrived at the following general picture of the phase transitions. At $T>T_+$ the smectic-A phase is realized.
Then the chemical potential $\mu$ is determined by the condition $\partial \Omega_A/\partial \mu=-N$, where $N={\cal N}/A$ is the average two-dimensional density of molecules of the film. At $T<T_-$ the hexatic phase is realized. Then the chemical potential $\mu$ is determined
by the condition $\partial \Omega_H/\partial \mu=-N$. At $T_-<T<T_+$ the phases coexist, then the chemical potential $\mu$ is determined
by the condition (\ref{ther3}). Thus, the chemical potential $\mu_+$ at $T=T_+$ is determined by the relation (\ref{ther3}) and the condition $\partial \Omega_A/\partial \mu=-N$.

Below $T_+$, in the region of the phase coexistence, the density of the smectic-A phase $N_A= -\partial\Omega_A/\partial\mu$
does not coincide with $N$. We expect that it is larger than $N$: $N_H<N<N_A$. Having in mind narrowness of the coexistence region,
we expand the density of the smectic-A phase in $\mu-\mu_+$, $T-T_+$ to obtain
 \begin{equation}
 N_A=N-\frac{\partial^2 \Omega_A}{\partial\mu^2}(\mu-\mu_+)
 -\frac{\partial^2 \Omega_A}{\partial T\partial\mu} (T-T_+),
 \label{ther11}
 \end{equation}
where the derivatives are taken at $T=T_+$, $\mu=\mu_+$.

\subsection{Landau expansion}
\label{sec:landau}

The hexatic order parameter $\psi$ (see its definition in \cite{CL00,GP93} and its symmetry derivation in \cite{GKL91}) in the region
of the phase coexistence is assumed to be small. Then one may expand the grand thermodynamic potential $\Omega$ in $\psi$
to obtain $\Omega=\Omega_0+\Phi$ where $\Phi$ is the Landau functional. In the context of the bulk consideration
(neglecting surface effects) the first terms of its expansion in $\psi$ are
 \begin{eqnarray}
 \Phi= L \left(a \,\vert \psi \vert^2 \,-\,\lambda\, \vert \psi \vert^4/6
 \,+\,\zeta \vert \psi \vert^6/90 \right),
 \label{ther4}
 \end{eqnarray}
where $L$ is the thickness of the film and the coefficients $a,\lambda,\zeta$ are functions of $T,\mu$.

We expanded the grand thermodynamic potential $\Omega$ up to the sixth order in $\psi$ having in mind that both coefficients, $a$ and
$\lambda$, are anomalously small. By other words, we are in the tricritical regime (near a tricritical point in the phase diagram).
It is well known \cite{LL80,ST87,AN91} that in the tricritical regime fluctuations of the order parameter are relatively weak:
they produce only logarithmic corrections to observable quantities. Therefore our problem can be examined in the mean field approximation.

To find equilibrium values of the order parameter $\psi$, one should minimize the Landau functional (\ref{ther4}). The smectic-A phase
corresponds to zero value of the order parameter $\psi$. The minimum of $\Phi$ at $\psi=0$ is realized if $a>0$, the condition is implied below.
The hexatic phase corresponds to a non-zero order parameter, that can be found as a result of the minimization:
 \begin{equation}
 | \psi_m |^2  =\frac{5\lambda}{\zeta}\left(1
 +\sqrt{1-\frac{6a\zeta}{5\lambda^2}}\right).
 \label{ther5}
 \end{equation}
This minimum of the Landau functional exists if $6a\zeta<5\lambda^2$.

In the mean field approximation the Landau functional $\Phi$ is equal to zero for the smectic-A phase.
Therefore $\Omega_A=\Omega_0$, $\Omega_H=\Omega_0+\Phi_H$. Since $\Omega_H=\Omega_A +\Phi_H$, we obtain
 \begin{equation}
 N_H=N_A-\partial\Phi_H/\partial\mu,
 \label{jump}
 \end{equation}
in the region of the phase coexistence. Note that at calculating the derivative in Eq. (\ref{jump}) one can differentiate solely
the coefficients in the expansion (\ref{ther4}) since $\partial\Phi/\partial\psi=0$ in the minimum.

To find the value of the order parameter in the regime of coexistence of the phases one should use the relation (\ref{ther3}).
In our case it leads to $\Phi_H=0$. Substituting the expression (\ref{ther5}) into Eq. (\ref{ther4}) and equating the result
to zero, one finds $a= a_0$, $|\psi|=\psi_0$, where the equilibrium value of the order parameter is
 \begin{eqnarray}
 \psi_0^2=15\lambda/(2\zeta), \quad  a_0= (5/8) \lambda^2/\zeta.
 \label{ther7}
 \end{eqnarray}
Thus, both parameters, $\psi$ and $a$, are fixed by the equilibrium conditions. Note the relation ${\zeta a}\sim{\lambda^2}$
between two small parameters, $a$ and $\lambda$.

Within Landau theory, the parameter $a$ in the expansion (\ref{ther4}) is the most sensitive to variations
of chemical potential $\mu$ and of temperature $T$. Therefore in the main approximation we can safely assume that
the coefficients $\lambda$ and $\zeta$ are independent of the temperature and the chemical potential in the phase coexistence region.
In the same spirit we believe that the equilibrium phase coexistence exists in the narrow range of the parameters governing the transition.
As we will show below it is the case in the vicinity of the tricritical point. Thus we expand $a$ in $T-T_+$ and $\mu-\mu_+$ to obtain
 \begin{equation}
 a=a_++\alpha(T-T_+) +\beta (\mu-\mu_+),
 \label{ther10}
 \end{equation}
where $a_+$ is the value of the parameter $a$ at $T=T_+$ and $\mu=\mu_+$. One expects that both parameters, $\alpha$ and $\beta$,
are positive. The conditions mean that at diminishing $T$ or $\mu$ the hexatic phase becomes more preferable.

In our model the only quantity in the Landau functional (\ref{ther4}), dependent on $\mu$, is $a$. Calculating $\partial\Phi/\partial\mu$,
and substituting then the value (\ref{ther7}), we find in accordance with Eq. (\ref{jump})
 \begin{equation}
 N_H=N_A-L\beta \frac{15\lambda}{2\zeta}.
 \label{ther12}
 \end{equation}
As we expected, there is an additional negative contribution to $N_H$ in comparison with $N_A$. In our model, it is independent of $T$.

The condition $a=a_0$ shows that at the phase coexistence $a$ remains approximately constant that is $\alpha(T-T_+)+\beta(\mu-\mu_+)=0$.
 Substituting the relation to the expression (\ref{ther11}) and resulting formula for $N_A$ to the expression (\ref{ther12}), one obtains
 \begin{eqnarray}
 N_H=N-L\beta \frac{15\lambda}{2\zeta}
 +\Gamma L(T_+-T),
 \label{ther13} \\
 \Gamma L=\frac{\partial^2 \Omega_0}{\partial T \partial\mu}
 -\frac{\alpha}{\beta}\frac{\partial^2 \Omega_0}{\partial\mu^2}.
 \label{ther9}
 \end{eqnarray}
The lower coexistence temperature $T_-$ is achieved where $N_H$ becomes $N$, the property enables one to obtain the temperature interval
of the phase coexistence  in bulk:
 \begin{equation}
 \frac{\Gamma}{\beta} (T_+-T_-)=\psi_0^2=
 \frac{15\lambda}{2\zeta}.
 \label{ther14}
 \end{equation}
Since the phase transition occurs in the vicinity of the tricritical point, the coefficient $\lambda $ is small.
Therefore the interval $T_+-T_-$ is also small, as we have assumed expanding the coefficient $a$ in (\ref{ther10}),  and
$a_+=a_-=a_0$ in the first order of the expansion  of $a$ in deviations $(T-T_+)$, $(\mu-\mu_+)$.

\section{Surface effects}
\label{sec:surface}

Here we consider effects related to the surface hexatic order (see, original publications \cite{CJ98,SG92,GS93},
and monograph \cite{PP06}, containing also many useful references). We assume that at the surface of the film the hexatic
order parameter is fixed and  $\psi_s$ is the absolute value of the hexatic order parameter $\psi$ at the surface.
Then the order parameter is   non-zero and inhomogeneous in space in the both phases.   In the spirit of the mean field treatment
we assume that $\psi$ is homogeneous along the film. However, due to the prescribed value of  the surface ordering,
 it is inhomogeneous in the orthogonal direction. To analyze the situation one should introduce the Landau functional
 for the inhomogeneous order parameter. For the purpose we add the gradient term to the Landau expansion (\ref{ther4}) and obtain
 \begin{eqnarray}
 \Phi=  \int_{-L/2}^{L/2}\!\! dz
 \left(b |\partial_z \psi|^2+a  |\psi|^2
 - \frac{\lambda|\psi|^4}{6}
 +\frac{\zeta|\psi|^6}{90} \right),
 \label{surf1}
 \end{eqnarray}
where $b$ is  Landau theory expansion coefficient and $z$ axis is along the smectic layer normal.

We place the plane $z=0$ in the middle of the film. Let us stress that the surface ordering provides a non-zero value of the Landau functional
for the smectic-A phase, in contrast to the analysis of the section \ref{sec:bulk}, performed neglecting surface effects. Since the gradient term
is positive, the homogeneous configuration is a trivial minimizer of the Landau thermodynamic potential. In the bulk system if the thermodynamic
potential is convex, a single homogeneous phase is a solution corresponding to a stable thermodynamic state. However if on the other hand
it is concave for some values of the model parameters, it is energetically favorable to split the system into (at least)
two regions with the phase coexistence. Conventional wisdom suggests that surface ordering plays a little role
for bulk transitions for sufficiently thick films. Although conventional wisdom is simple and comfortable but not necessary
always true. We will show in this section, that it is just the case for $Sm A$ -- $Hex B$ transition in the vicinity of the tricritical point.

The characteristic length of the order parameter variations is its correlation length $\xi$, defined as
 \begin{equation}
 \xi^2=\frac{b}{a_0}=\frac{8b\zeta}{5\lambda^2}.
 \label{corrl}
 \end{equation}
The quantity $\xi$ is assumed to be much larger than the molecular length, the property holds because the system is assumed to be close
to a tricritical point. That justifies our phenomenological approach. It is worth to noting that in our approach the correlation length $\xi$
weakly depends on temperature in the coexistence region.

Further on we assume that the order parameter $\psi$ is real. The case corresponds to the minimum of the contribution
to the gradient term in the Landau expansion (\ref{surf1}), related to the gradient of the phase of the order parameter $\psi$.
Based on symmetry reasoning, we consider the symmetric in $z$ profile of the order parameter: $\psi$ is equal to $\psi_s$
at $z=\pm L/2$ and achieves a minimum at $z=0$.

Varying the Landau functional (\ref{surf1}) over $\psi$, one finds the extremum condition
 \begin{eqnarray}
 -b \partial_z^2 \psi+a  \psi
 -\lambda \psi^3/3
  +\zeta \psi^5/30 =0.
 \label{surf2}
 \end{eqnarray}
The equation (\ref{surf2}) has the first integral:
 \begin{eqnarray}
 -(\partial_z\psi)^2 + g(\psi)=\gamma, \qquad
 \label{surf3} \\
 g(\psi)=\frac{1}{b}\left( a  \psi^2
 -\lambda \psi^4/6  +\zeta \psi^6/90 \right),
 \label{surf28}
 \end{eqnarray}
where $\gamma$ is an $z$-independent parameter. As it follows from Eq. (\ref{surf3}), $\gamma$ is the value of $g$
at $z=0$ where $\partial_z\psi=0$ (since $\psi$ is symmetric in $z$).

With the relation (\ref{surf3}) taken into account, the energy (\ref{surf1}) becomes
 \begin{equation}
 \Phi= 2b \int_{0}^{L/2} dz \
 (2g -\gamma).
 \label{surf5}
 \end{equation}
The equation (\ref{surf3}) at $z>0$ is rewritten as $\partial_z \psi=\sqrt{g-\gamma}$, that is $dz=d\psi/\sqrt{g-\gamma}$.
Integrating the condition, we find
 \begin{equation}
 L=2\int_{\psi_\star}^{\psi_s}
 \frac{d\psi}{\sqrt{g-\gamma}},
 \label{surf6}
 \end{equation}
where $\psi_\star$ is the value of the order parameter at $z=0$, $\gamma=g(\psi_\star)$ in accordance with Eq. (\ref{surf3})
and $\psi_s$ is the surface value of the order parameter.

Analogously, the Landau functional (\ref{surf5}) can be rewritten as
 \begin{equation}
 \Phi=2b \left[2 \int_{\psi_\star}^{\psi_s}
 d\psi\  \sqrt{g-\gamma} +\gamma L/2\right].
 \label{surf7}
 \end{equation}
The expression determines the smectic energy per unit area of the phase with the surface conditions taken into account.
Note that the relation (\ref{surf6}) can be treated as the extremum condition in terms of $\psi_\star$ (or $\gamma$)
of the Landau functional (\ref{surf7}).

In the vicinity of the tricritical point $\psi_s$ entering  into Eqs. (\ref{surf6},\ref{surf7}), is much larger
than the characteristic values of the order parameter $\psi$ in  the bulk. Therefore one can put $\psi_s\to\infty$
in Eq. (\ref{surf6}) due to convergence of the integral. Thus we arrive at the function
 \begin{equation}
 \Xi(a,\gamma)=2 \int_{\psi_\star}^\infty \frac{d\psi}{\sqrt{g-\gamma}},
 \label{surf32}
 \end{equation}
to be equated to $L$ in equilibrium in accordance with Eq. (\ref{surf6}).

\subsection{Phase coexistence}

In the phase coexistence region there are two different solutions of the equation (\ref{surf3})
satisfying the conditions (\ref{surf6}) and corresponding to the same energies, $\Phi_A=\Phi_H$.
We designate as $\psi_A$, $\psi_H$ the values of the order parameter at $z=0$ in the $Sm A$ phase
and in the $Hex B$ phase, respectively. Introducing also $\gamma_A=g(\psi_A)$ and $\gamma_H=g(\psi_H)$, we arrive at the relations
 \begin{equation}
 \Xi(a,\gamma_A)=L=\Xi(a,\gamma_H).
 \label{surf8}
 \end{equation}
The relations (\ref{surf8}) together with the condition $\Phi_A=\Phi_H$ are three equations for the three variables $L,\gamma_A,\gamma_H$.

Now we are in the position to find the difference $\Delta\Phi=\Phi_A-\Phi_H$:
 \begin{eqnarray}
 \frac{\Delta\Phi}{b}=(\gamma_A-\gamma_H)L
 + 4\int _{\psi_A}^{\psi_H} d\psi \sqrt{g-\gamma_A}
 \nonumber \\
 +4\int_{\psi_H}^\infty d\psi
 \left(\sqrt{g-\gamma_A}-\sqrt{g-\gamma_H}\right),
 \label{surf9}
 \end{eqnarray}
in accordance with Eq.  (\ref{surf7}). Again, we extended the integration up to infinity due to convergence of the integral. One can easily
check that
 \begin{equation}
 \frac{\Delta\Phi}{b}
 =\int_{\gamma_A}^{\gamma_H} d\gamma (\Xi-L).
 \label{surf33}
 \end{equation}
Note that the equations (\ref{surf8}) are extrema conditions for the quantity (\ref{surf33}) in terms of $\gamma_A$ and $\gamma_H$.

The difference $\Delta\Phi$ can be considered as a function of $L$,
with the relation $b^{-1}\partial(\Delta\Phi) /\partial L= \gamma_A -\gamma_H<0$. In addition, $\Delta\Phi$
is a function of $a$ via the function $g$, see Eq. (\ref{surf28}). Then the equilibrium condition $\Delta\Phi=0$
determines $L$ as a function of $a$. Therefore one obtains
 \begin{equation}
 \frac{\partial\Delta\Phi}{\partial a}
 +b(\gamma_A-\gamma_H) \frac{dL}{da}=0.
 \label{deriv}
 \end{equation}
Since the relations (\ref{surf8}) are extrema conditions of $\Delta\Phi$ in terms of $\psi_A,\psi_H$, one finds
 \begin{eqnarray}
 \frac{\partial\Delta \Phi}{\partial a}=
 2\int _{\psi_A}^{\psi_H} d\psi\, \psi^2/ \sqrt{g-\gamma_A}
 \nonumber \\
 +2\int_{\psi_H}^\infty d\psi\, \psi^2
 \left(1/\sqrt{g-\gamma_A}-1/\sqrt{g-\gamma_H}\right).
 \label{surf22}
 \end{eqnarray}

\begin{figure}
    \includegraphics[width=1\linewidth]{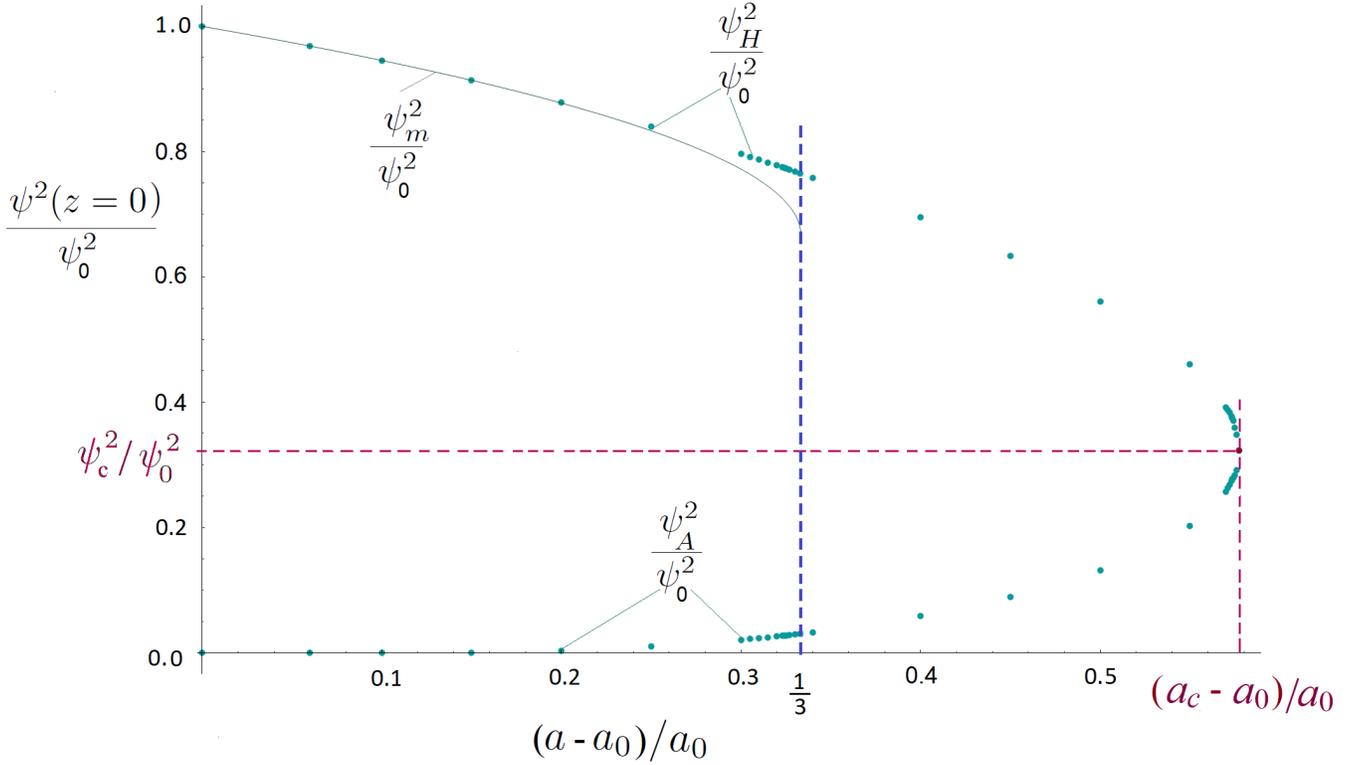}
    \caption{
Found numerically by solving Eqs. (\ref{surf23},\ref{surf25}) values of the dimensionless value of the hexatic
order parameter $\psi^2/\psi_m^2\,$ at $z=0$ in the coexistence regime (sea-green circles) as a function of the
dimensionless temperature deviation $(a-a_0)/a_0$. The dark green solid line shows corresponding analytic result
(Eq. (\ref{ther5})), and the blue vertical dash line shows the limit of the existence for this local minimum ($(a-a_0)/a_0\le\,1/3$).
             \label{psi}
              }
              \end{figure}

To find the value of $L$ at a given $a$, one should solve the system of equations (\ref{surf8}) together
 with the condition $\Delta\Phi=0$. The relations are reduced to the system of equations
 \begin{eqnarray}
 \Xi(\gamma_A,a)=\Xi(\gamma_H,a), \qquad
 \label{surf23} \\
 \int _{\psi_A}^{\psi_H} d\psi\, (2g-\gamma_A)/ \sqrt{g-\gamma_A} \qquad
 \nonumber \\
 +\int_{\psi_H}^\infty d\psi\,
 \left[\frac{2g-\gamma_A}{\sqrt{g-\gamma_A}}
 -\frac{2g-\gamma_H}{\sqrt{g-\gamma_H}}\right]=0,
 \label{surf25}
 \end{eqnarray}
determining $\psi_A,\psi_H$ (see   Fig. \ref{psi}). The function $\Xi$ in Eq. (\ref{surf23}) is defined by Eq. (\ref{surf32}).

\begin{figure}
    \includegraphics[width=1\linewidth]{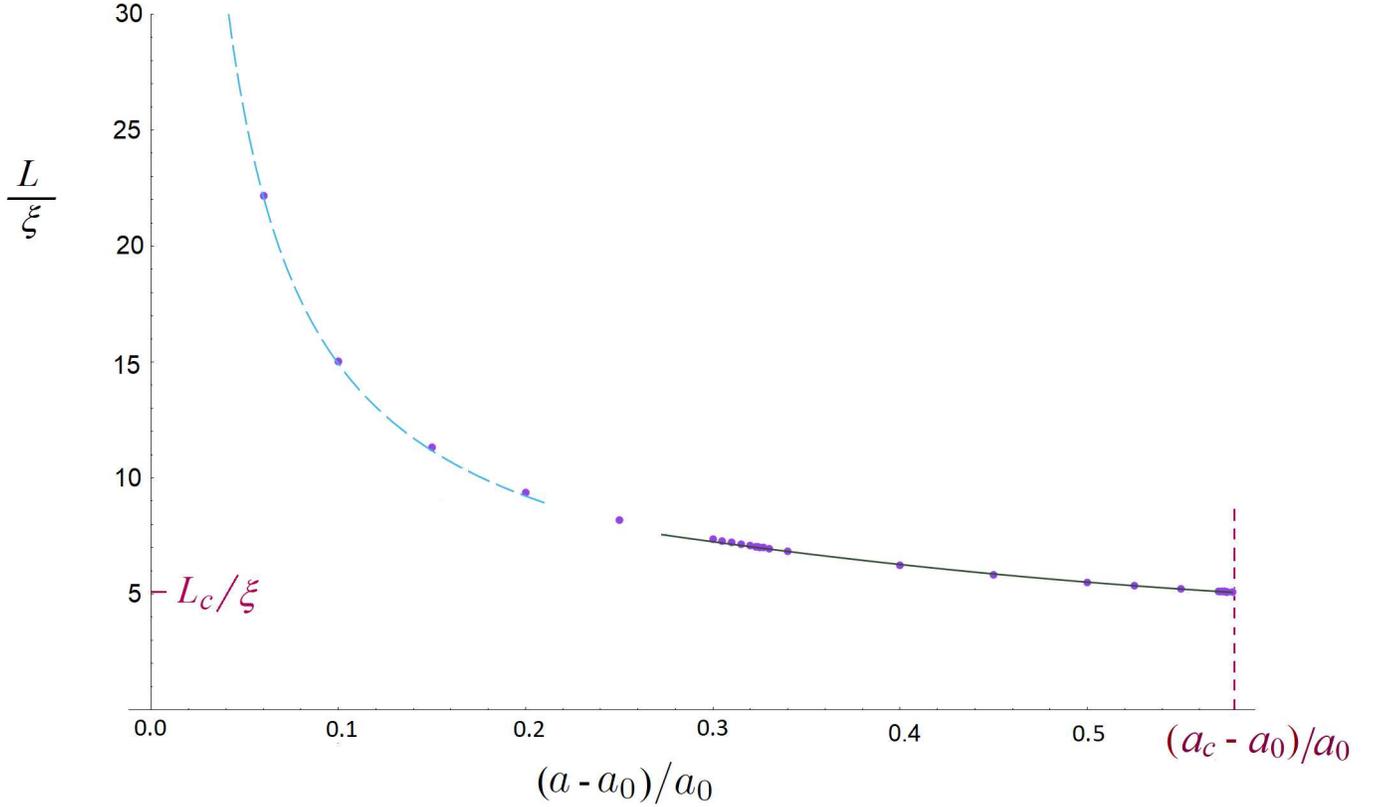}
    \caption{
Found numerically by solving Eqs. (\ref{surf23},\ref{surf25},\ref{surf8}) values of the dimensionless film thickness
$L/\xi\,$  in the coexistence regime (lilac circles) as a function of the dimensionless temperature deviation  $(a-a_0)/a_0$. The
azure dash line  in the left shows corresponding analytic result (Eq. (\ref{differ9})).  Green solid line (for $(a-a_0)/a_0<\,(a_c-a_0)/a_0$)
is obtained by fitting with Eq. (\ref{thin4}) the result  near the point $\gamma=g_c$.
             \label{Ltda}
              }
              \end{figure}

\begin{figure}
    \includegraphics[width=1\linewidth]{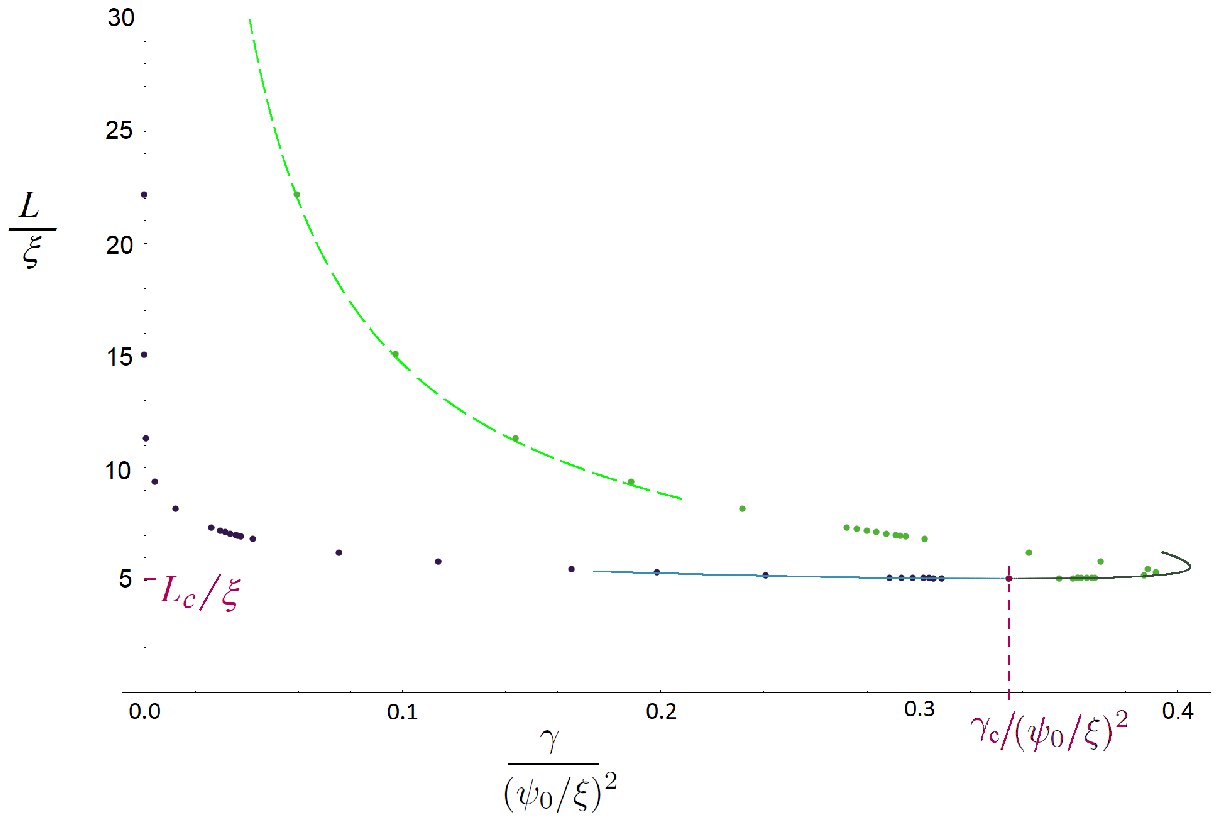}
    \caption{
Numerical values (found by solving Eqs. (\ref{surf23},\ref{surf25}), with Eq. (\ref{surf8}) and using Eqs.
(\ref{surf28},\ref{surf4})) for the dimensionless film thickness $L/\xi\,$ in the coexistence regime of  $Sm A$  (dark blue
circles) and $Hex B$ phase (green circles) presented as a function of the dimensionless variable $\gamma/(\psi_0/\xi)^2$. The last
right numerical point corresponds to the condition $\partial\gamma/\partial L\,=0$. Corresponding analytical
description according to the Eq. (\ref{differ9}), using Eq. (\ref{surf4}), is shown by the light green dash line in the left.
Fitting, see Eqs. (\ref{thin4},\ref{thin5}) in the vicinity of the point $\gamma=g_c$ is shown by blue solid line (for the $Sm A$
phase, $\gamma\le\,\gamma_c$) and by the dark green line (for the $Hex B$ phase, $\gamma\ge\,\gamma_c$).
             \label{Lg}
              }
              \end{figure}

After solving the system of equations (\ref{surf23},\ref{surf25}), $L$ can be found from one of the relations (\ref{surf8}).
By other words, $L$ is determined by the relation $L=\Xi(\gamma_A,a)$. The results of the corresponding numerical calculations
are shown in Figs. \ref{Ltda}, \ref{Lg}. Here and below numerical solutions of the system of the equations
(\ref{surf23},\ref{surf25},\ref{surf12}) were obtained. For this purpose Wolfram Mathematica Professional Version Premier
Service L3159-1472 was used. The numeric errors of all dimensionless solving of the discussed equations is less than $5\cdot\,10^{-\,6}$.

\subsection{Model Landau functional}

Here we exploit the model, introduced above, where the parameter $a$ is determined by the expansion (\ref{ther10})
and the parameters $\lambda,\zeta$ are treated as constants, independent of temperature and chemical potential.
In addition, we assume that the parameter $b$ is constant as well. Then one finds from Eqs. (\ref{surf1},\ref{surf3})
 \begin{equation}
 N_A=-\frac{\partial\Omega_A}{\partial\mu}
 =-\frac{\partial\Omega_0}{\partial\mu}
 -2\beta \int_{\psi_A}^{\psi_s}
 \frac{d\psi \ \psi^2}{\sqrt{g-\gamma_A}},
 \label{surf10}
 \end{equation}
and an analogous expression for the hexatic phase.

Since $N_A=N$ at $T=T_+$ and $N_H=N$ at $T=T_-$, then the interval of the phase coexistence is determined
by the condition $N_H(T_-) =N_A(T_+)$. According to Eq. (\ref{surf10}), it is written as
 \begin{eqnarray}
 \frac{\partial\Omega_0}{\partial\mu}(T_-,\mu_-)
 -\frac{\partial\Omega_0}{\partial\mu}(T_+,\mu_+)
 = 2\beta \int_{\psi_A}^{\psi_H}
 \frac{d\psi \ \psi^2}{\sqrt{g_+-\gamma_A}}
 \nonumber \\
 +2\beta \int_{\psi_H}^\infty
 d\psi \ \psi^2 \left(\frac{1}{\sqrt{g_+-\gamma_A}}
 -\frac{1}{\sqrt{g_--\gamma_H}}\right), \qquad
 \label{surf11}
 \end{eqnarray}
where we, again, extended the integration up to infinity due to convergence of the integral. Here the parameters in $g_+$ and $\psi_A$ are
taken at $T=T_+$ and the parameters in $g_-$ and $\psi_H$ are taken at $T=T_-$.

\begin{figure}
    \includegraphics[width=1\linewidth]{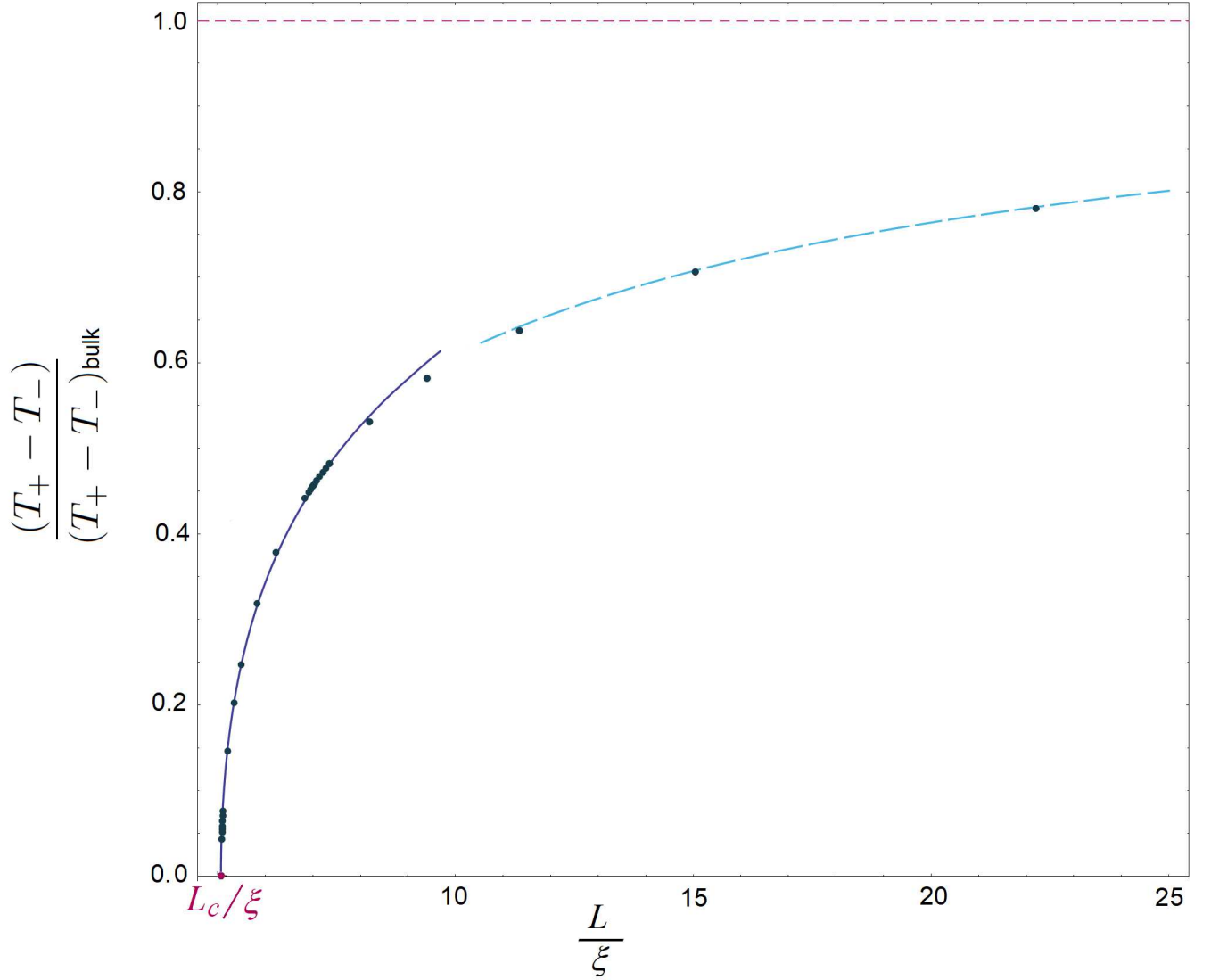}
    \caption{
The temperature interval of the phase coexistence as a function of the dimensionless film thickness $L/\xi\,$,
computed by solving numerically   Eqs. (\ref{surf23}),  (\ref{surf25}), (\ref{surf12}) (dark blue points).
Our theory result for the thick films (Eq. (\ref{differ103}) is shown by the azure dash line in the right.
Violet solid line (started from the point $L_c/\xi$) stands for the analytic
solution to the Eqs. (\ref{interval},\ref{surf40},\ref{surf41},\ref{thin4}) in a vicinity of the point $a=a_c$.
             \label{FigureDT}
    }
\end{figure}

In our model $a$ is $L$-dependent parameter in the coexistence region. Therefore we obtain from the expression (\ref{ther10})
$\alpha(T-T_+) +\beta (\mu-\mu_+)=0$. This means $\,g_{+}=g_{-}=g\,$, that is $a_+=a_-=a$ in the first order of the expansion  of $a$
in the deviations $(T-T_+)$, $(\mu-\mu_+)$. In this way expanding the difference of the derivatives in Eq. (\ref{surf11})
in $T_+-T_-$, $\mu_+-\mu_-$ and using the condition, we find the relation
 \begin{eqnarray}
 \frac{L\Gamma}{2\beta}(T_+-T_-) +
 \int_{\psi_A}^{\psi_H}
 \frac{d\psi \ \psi^2}{\sqrt{g-\gamma_A}}
 \nonumber \\
 + \int_{\psi_H}^\infty
 d\psi \ \psi^2 \left(\frac{1}{\sqrt{g-\gamma_A}}
 -\frac{1}{\sqrt{g-\gamma_H}}\right)=0,
 \label{surf12}
 \end{eqnarray}
which determines the interval of the phase coexistence, see Fig.  \ref{FigureDT}. The relation (\ref{surf12}) can be rewritten as
 \begin{equation}
  \frac{L\Gamma}{\beta}(T_+-T_-) +
  \frac{\partial\Delta\Phi}{\partial a}=0,
  \label{interval}
  \end{equation}
as a consequence of Eq. (\ref{surf22}). These relations (\ref{surf12},\ref{interval}) are our main results in the work, and they are ready
for further experimental inspection and theoretical analysis.

\subsection{Universal phase diagram}

To get further insight into the nature of the equilibrium phase coexistence, it is convenient to utilize the dimensionless
variables $a/a_0$ and $L/\xi$. Then we obtain a universal picture, independent of the concrete values of the model parameters, from
the results of the previous subsection. Particularly, one can relate the variables $a/a_0$  and $L/\xi$. Although the solution
of the above non-linear equations can be found only numerically, we one can formulate some general universal laws valid (within our
model assumptions) in the equilibrium phase coexistence region. Namely, the relations (\ref{surf8}) imply that in the coexistence
regime the equation $\Xi(a,\gamma)=L$ should have at least two solutions. Already this deceptively simple observation restricts
the values of our model parameters. Let us first look at the function $\Xi$.

    \begin{figure}
    \includegraphics[width=0.8\linewidth]{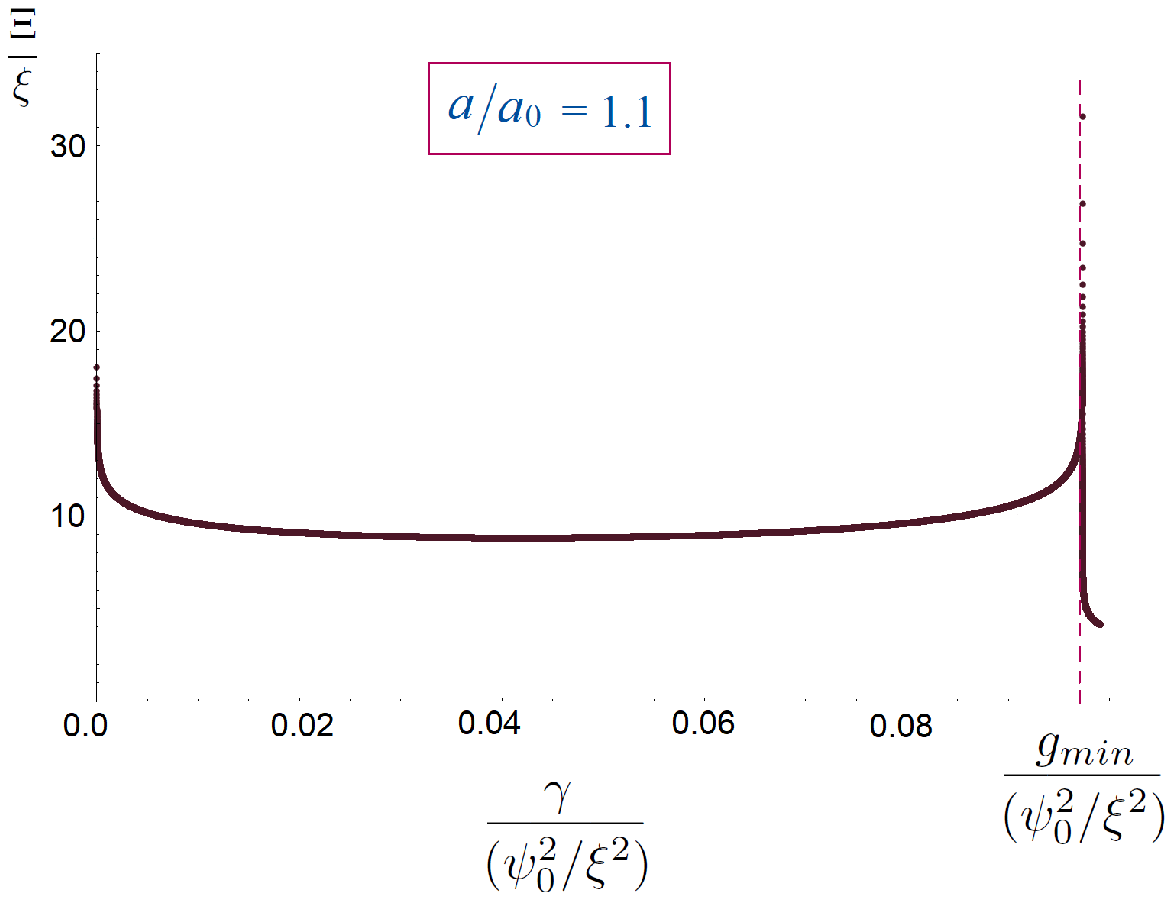}
    \caption{
Computed from Eq.(\ref{surf32}) for $a/a_0=1.1$ (i.e. for the case $1<a/a_0<4/3$) function  $\Xi/\xi$ versus $\gamma/(\psi_0/\xi)^2$.
                  \label{FigureDT0}
    }
 \end{figure}

\begin{figure}
    \includegraphics[width=0.4\linewidth]{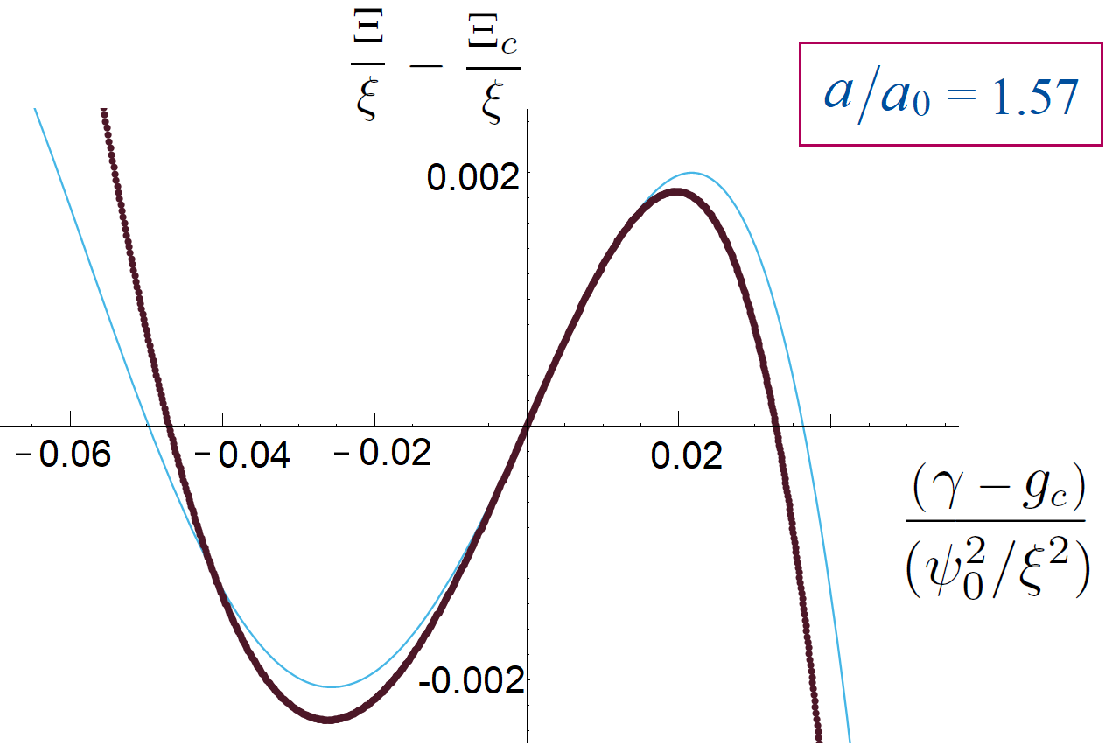}
     \includegraphics[width=0.4\linewidth]{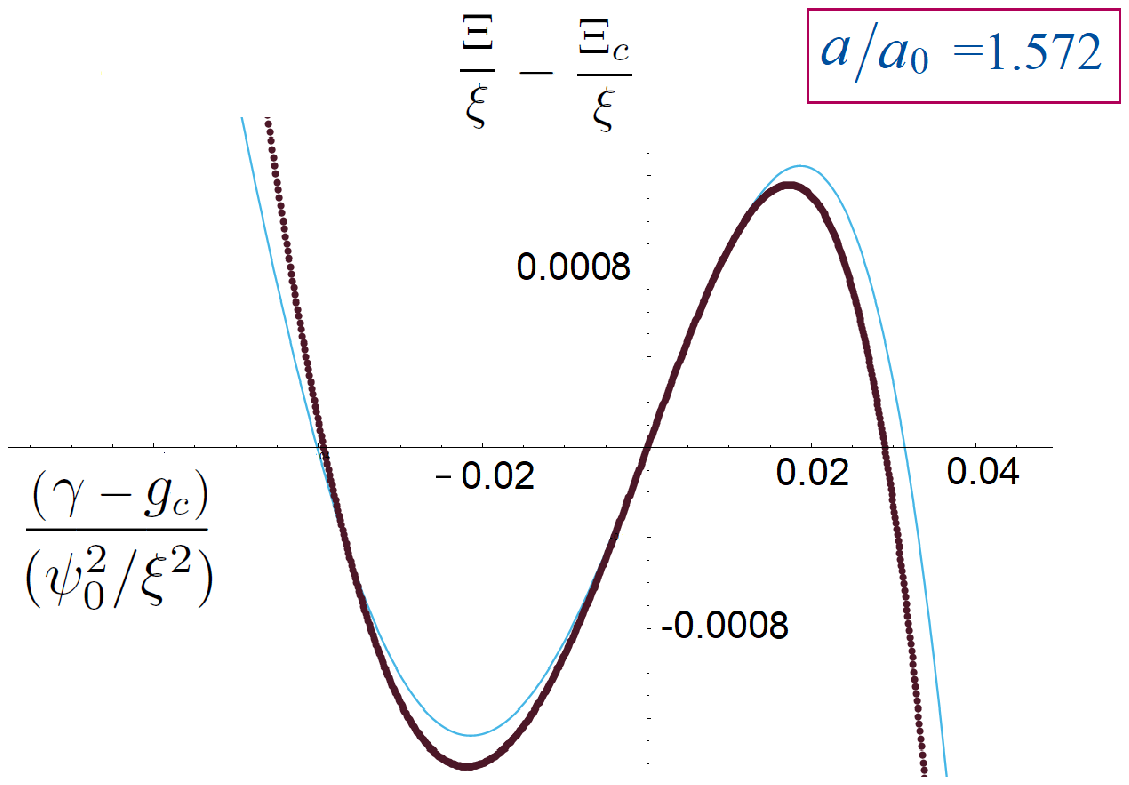}
    \caption{
Comparison the numeric solution of Eq.(\ref{surf32}) for ($\Xi/\xi$ as a function of $(\gamma - g_c)/(\psi_0/\xi)^2$) )
with our theory analysis (Eq.(\ref{thin1})) (thin azure solid line). $a/a_0=1.57$  and $a/a_0=1.572$ (i.e. for the case $a/a_0 >4/3$).
                  \label{FigureDT1}
    }
\end{figure}

\begin{figure}
    \includegraphics[width=0.45\linewidth]{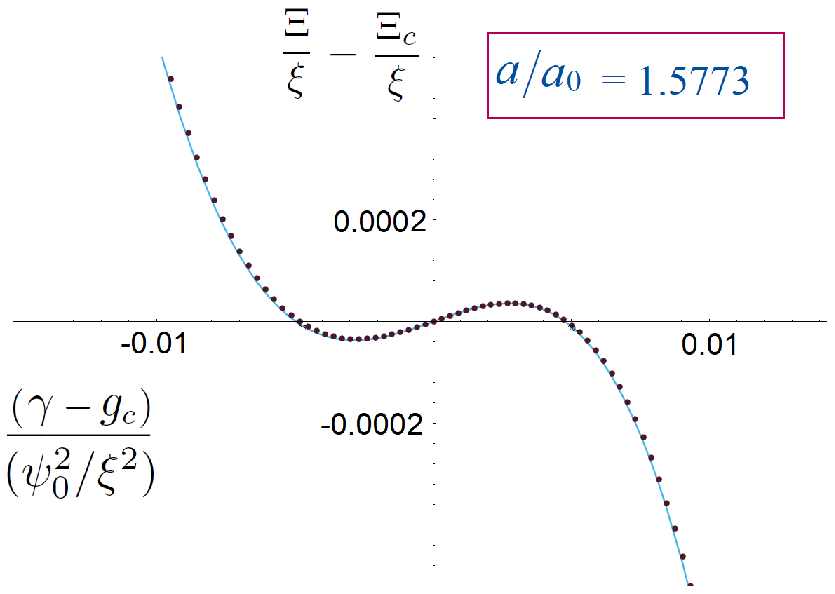}
     \includegraphics[width=0.45\linewidth]{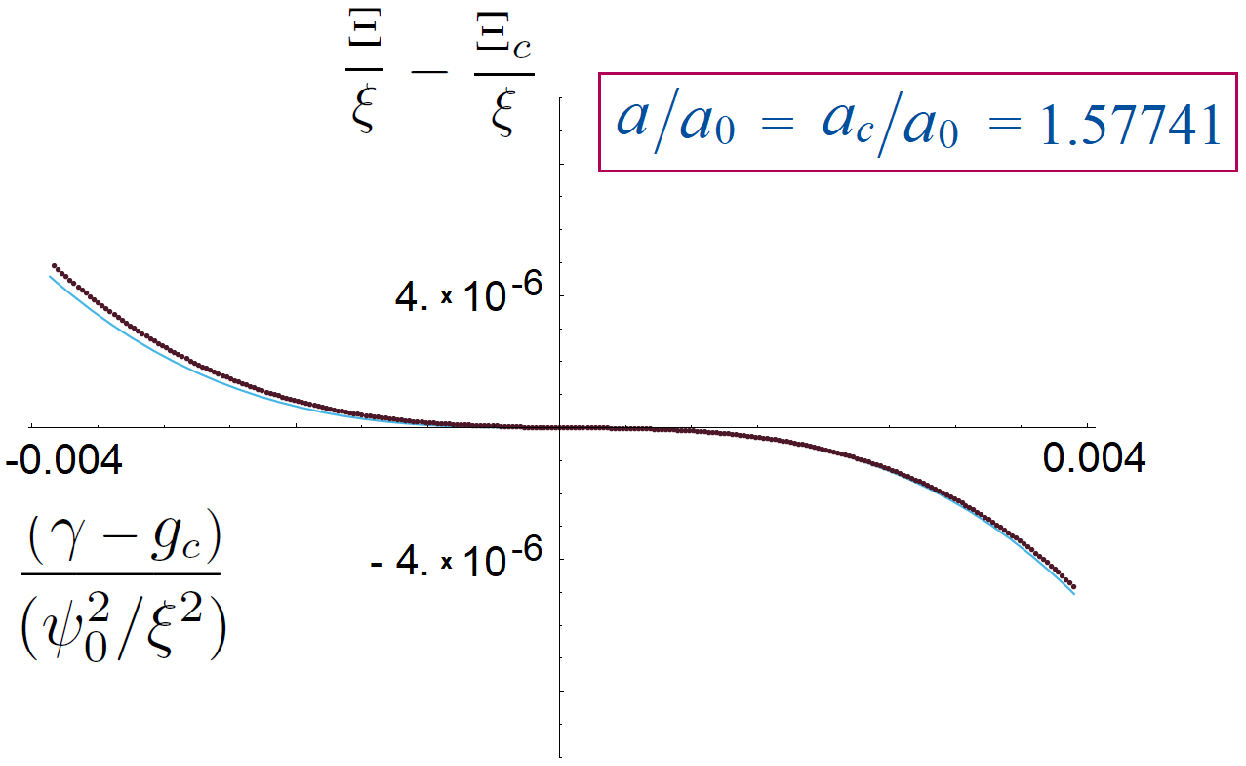}
    \caption{
Comparison of numerically found by solving Eq. (\ref{surf32}) $\Xi/\xi$ as a function of $(\gamma - g_c)/(\psi_0/\xi)^2$
with analytical approximation Eq. (\ref{thin1}) (thin azure solid line). $a/a_0=1.5773$ (i.e. for the case $a/a_0 > 4/3$)
and $a/a_0=a_c/a_0=1.57741$.
                  \label{FigureDT2}
    }
\end{figure}

For small $\gamma$ the function $\Xi$ diverges logarithmically, see Fig. \ref{FigureDT0}. If $1<a/a_0<4/3$, then the function
$g(\psi)$ (\ref{surf28}) has a minimum at non-zero $\psi$. Therefore the function $\Xi$ logarithmically diverges at
$\gamma\to g_{min}$ where $g_{min}$ is the minimal value of the function $g$. Thus $\Xi$ has a minimum inside the interval
$0<\gamma<g_{min}$, see Fig. \ref{FigureDT0}. At $\gamma>g_{min}$ the function $\Xi$ monotonously decreases as $\gamma$ grows. At
$a/a_0>4/3$ the minimum in the function $g(\psi)$ disappears and $\Xi$ becomes a regular function of $\gamma$. However, $\Xi$
remains a non-monotonic function of $\gamma$ (it has a minimum and a maximum, see Figs. \ref{FigureDT1}-\ref{FigureDT2} up to some
critical value $a_c$, the value of $a_c/a_0$ is approximately equal to $1.5774$, see Fig. \ref{FigureDT2}. At $a>a_c$ the function
$\Xi(\gamma)$ becomes monotonic.

\begin{figure}
    \includegraphics[width=1\linewidth]{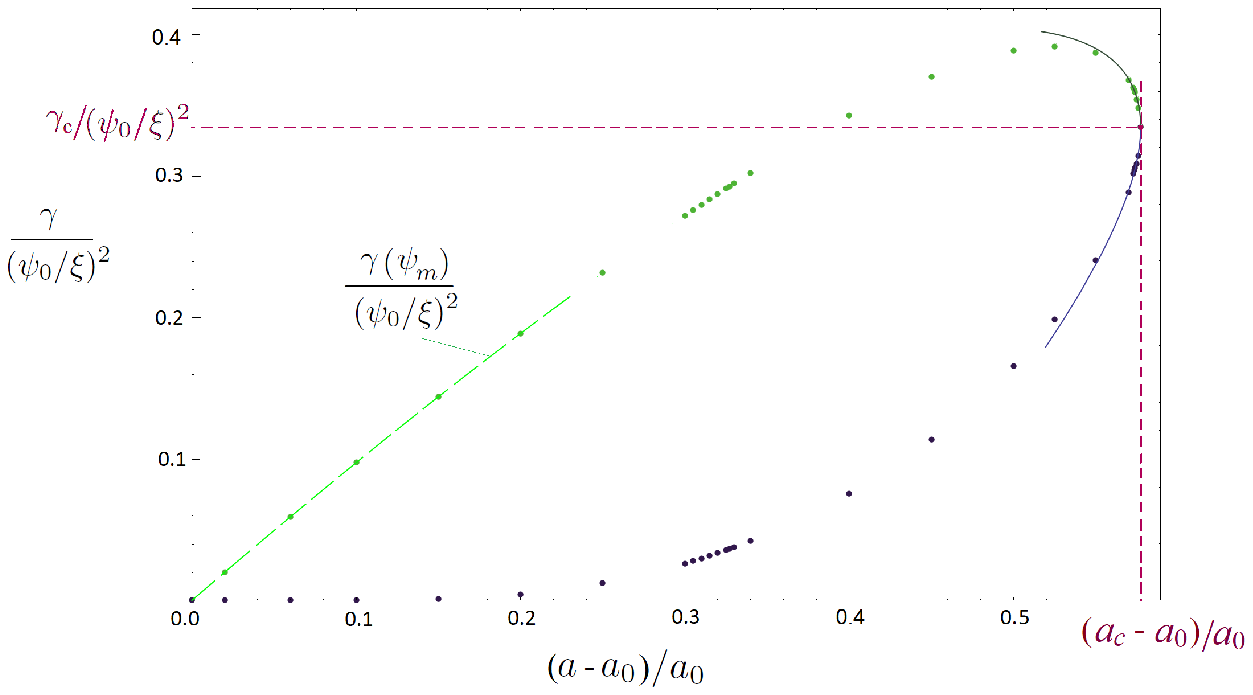}
    \caption{
Numerically found by solving Eqs. (\ref{surf23},\ref{surf25}), using Eq. (\ref{surf28}, \ref{surf4}), values of $\gamma_A$ and
$\gamma_H$  for the $Sm A$ (blue circles) and $Hex B$ phase (green circles) as a function of  $(a-a_0)/a_0$. Light green dash line
represents the analytical dependence $\gamma(\psi_m)/(\psi_0\,\xi^{-1})^2$. Using Eqs. (\ref{thin5}) we fit the
numerical data near the point $\gamma=g_c$. Solid lines dark blue for $Sm A$ ($\gamma\le\,\gamma_c$) and dark green for $Hex B$
phase ($\gamma\ge\,\gamma_c$).
             \label{FigureR}
    }
\end{figure}

Thus, at $a<a_c$ there are three solutions of the equation $\Xi=L$ in some interval of the film thickness $L$.
The smallest by the value of $\psi$ solution corresponds to the $Sm A$ phase, $\gamma=\gamma_A$.
The next by its $\psi $ value solution corresponds to an unstable state. And finally the third solution with the biggest
$\psi$ corresponds to the $Hex B$ phase, $\gamma= \gamma_H$. In the limit $a\to a_c$ we find $\gamma_A \to\gamma_H$, and at $a>a_c$
there remains the only one solution of the equation $\Xi(a,\gamma)=L$. Then the equilibrium phase coexistence region shrinks to zero.
This result which have been emanated from our analysis, states that the equilibrium phase coexistence
is possible only in the interval $a_0<a<a_c$. Different values of $a$ correspond to the different
values of the film thickness $L$, see Fig. \ref{FigureR}. Found numerically value of $L_c/\xi$ is equal to $5.07$.

\subsection{Thick films}

In this subsection we analyze the case of large film thickness, $L\gg \xi$. The limit has been discussed
at the semi-quantitatively level in \cite{ZK18}. Here we present the quantitative theory. For the thick films naturally
the deviations of the film properties from the bulk ones are relatively weak. Particularly, the value of the parameter
$a$ is close to its bulk value $a_0$, $a-a_0\ll a_0$. It follows from the relations (\ref{surf8}) that at $L\gg\xi$
the integral (\ref{surf32}) is anomalously large. It enables us to develop the consistent analytical procedure
to study the surface effects in the equilibrium phase coexistence regime.

Let us turn to the hexatic phase. The value of $\psi$ in the
hexatic phase, $\psi_H$, is close to $\psi_m$, that corresponds to
the minimum of $g$, see Eq. (\ref{ther5}). The main contribution
to the integral (\ref{surf32}) stems from the vicinity of
$\psi_m$. Near $\psi_m$ the function $g$ can be approximated as
 \begin{equation}
 g=\frac{\psi_0^2}{b}\left[a-a_0+\frac{\lambda}{3}(\psi-\psi_m)^2\right].
 \label{surf15}
 \end{equation}
Starting from Eq. (\ref{surf15}) and using Eqs. (\ref{surf32},\ref{surf8}), we find with the logarithmic accuracy
 \begin{eqnarray}
 \frac{L}{\xi}=\ln\frac{\psi_0}{\psi_H-\psi_m}.
 \label{surf16}
 \end{eqnarray}
Thus, $\psi_H-\psi_m$ is exponentially small in $L/\xi$.

Let us now turn to the $Sm A$ phase. At $L\gg \xi$ the main contribution to the integral (\ref{surf32}) comes from
the small $\psi$, where $g\approx a\psi^2/b$. Calculating the integral with the logarithmic accuracy, one obtains
 \begin{equation}
 {L}/{\xi}=2 \ln(\psi_0/\psi_A).
 \label{surf14}
 \end{equation}
We conclude from Eq. (\ref{surf14}), that $\psi_A$ is exponentially small in $L/\xi$.

Now we use the condition $\Delta \Phi=0$, see Eq. (\ref{surf9}), to find $a$ at a given $L$. We can substitute
into Eq. (\ref{surf9}) $\psi_A=0$ and $\psi_H=\psi_m$. In the main approximation one obtains
 \begin{equation}
 \frac{a_0}{a-a_0}+\ln \frac{a_0}{a-a_0} =\frac{L}{\xi}.
 \label{surf18}
 \end{equation}
We see, that $(a-a_0)/a_0$ is a power of $\xi/L$, that justifies the substitution $\psi_A\to0$ and $\psi_H- \psi_m\to0$
since the quantities are exponentially small.

Note that for the $Sm A$ phase there is an additional logarithmic contribution to the integral in Eq. (\ref{surf32}), related
to a vicinity of the minimum of $g(\psi)$, containing $\ln[a_0/(a-a_0)]$. As it follows from Eq. (\ref{surf18}), the logarithm is
$\ln(L/\xi)$. Therefore the contribution is irrelevant in comparison with $L/\xi$ in the left hand side of Eq. (\ref{surf14}).

Now we rewrite Eq. (\ref{surf12}) as
 \begin{eqnarray}
 \frac{L\Gamma}{2\beta}(T_+-T_-) +
 \int_{0}^{\psi_m}
 \frac{d\psi \ \psi^2}{\sqrt{g}}
 \nonumber \\
 + \int_{\psi_m}^\infty
 d\psi \ (\psi^2-\psi_m^2) \left(\frac{1}{\sqrt{g}}
 -\frac{1}{\sqrt{g-\gamma_H}}\right)
 \nonumber \\
 + \psi_m^2 \int_{\psi_m}^\infty
 \frac{d\psi}{\sqrt{g}}
 -\psi_m^2 \int_{\psi_H}^\infty
 \frac{d\psi}{\sqrt{g-\gamma_H}} =0,
 \label{surf19}
 \end{eqnarray}
where we substituted $\psi_A=0$, $\psi_H=\psi_m$. The last term in Eq. (\ref{surf19}) is equal to $\psi_m^2 L/2$,
in agreement with Eq. (\ref{surf8}).

In the main approximation we find
 \begin{equation}
 \frac{\Gamma}{\beta\psi_0^2}(T_+-T_-)=
 1-\frac{\xi}{L} \ln\frac{L}{\xi},
 \label{surf20}
 \end{equation}
where we used Eq. (\ref{surf18}). The expression (\ref{surf20}) gives the first correction to the bulk expression (\ref{ther14}).
The contributions leading to the logarithmic factor in the Eq. (\ref{surf20}) were missed in the work \cite{ZK18}. Therefore
the expression for the temperature width of the phase coexistence region presented in \cite{ZK18} can be used only for qualitative
interpretation of the data (note however that in terms of numeric values for the range of the film thicknesses considered in
\cite{ZK18}, the logarithmic factor is almost irrelevant). Nevertheless the
logarithmic factor is very important conceptually. Thanks to this factor we are in the position to perform consistently our calculations with
logarithmic accuracy (see, however, the appendices with higher order corrections included). Besides it allows us to distinguish found above
law for the temperature width of the coexistence region, from regular (existing in any system) finite size corrections which scales as
$\xi /L$. The fact that $\psi_A$ and $\psi_H-\psi_m$ are exponentially small, enables us to find analytically next terms of
the expansion in the parameter $\xi/L$ in the expression for $T_+-T_-$. The corresponding analysis is placed into Appendix
\ref{sec:thick}, see also Figs. \ref{Ltda}, \ref{Lg}.

\subsection{Thin films}

Being interested in thin films, we consider the case $a>4/3 a_0$. Then the quantity $\Xi$ (\ref{surf32}) has no singularities,
as a function of $\gamma$. However, at $a<a_c$ it is still a non-monotonic function of $\gamma$. At $a=a_c$ the function $\Xi$ (\ref{surf32})
has a point $\gamma=g_c$, where both, $\partial\Xi/\partial \gamma$ and $\partial^2\Xi/\partial \gamma^2$ are equal to zero.

In the vicinity of the point the quantity $\Xi$ can be approximated as
 \begin{eqnarray}
 \frac{\Xi}{\xi}=\frac{\Xi(a,y_c)}{\xi}
 -A(y-y_c)^3
 -B \frac{a-a_c}{a_0}(y-y_c),
 \label{surf34}
 \end{eqnarray}
where $\gamma=(\psi_0^2/\xi^2)y$ and $A,B$ are dimensionless constants. Their numerical values are $A=82.1362$, $B=17.6392$.
 Exploiting Eq. (\ref{surf33}), one finds from Eq. (\ref{surf34})
 \begin{eqnarray}
 \frac{\xi\Delta\Phi}{\psi_0^2}
 =\left[\frac{\Xi(a,y_c)}{\xi}-\frac{L}{\xi}\right](y_H-y_A)
 \nonumber \\
 -\frac{A}{4}\left[
 (y_H-y_c)^4-(y_A-y_c)^4 \right]
 \nonumber \\
 -\frac{B}{2} \frac{a-a_c}{a_0}
 \left[(y_H-y_c)^2-(y_A-y_c)^2 \right].
 \label{surf35}
 \end{eqnarray}

Now we can find the equilibrium values of the parameters that are determined by the conditions (\ref{surf8})
and $\Delta\Phi=0$. The conditions (\ref{surf8}) are written as
 \begin{eqnarray}
 \Xi(a,y_c)/\xi-L/\xi \qquad
 \nonumber \\
 =A(y_H-y_c)^3+B\frac{a-a_c}{a_0}(y_H-y_c)
 \nonumber \\
 =A(y_A-y_c)^3+B\frac{a-a_c}{a_0} (y_A-y_c).
 \label{surf37}
 \end{eqnarray}
Equating then $\Delta\Phi$ to zero, we find from Eqs. (\ref{surf35}) and (\ref{surf37})
 \begin{eqnarray}
 L=\Xi(a,y_c), \qquad
 \label{surf38} \\
 y_H-y_c=\sqrt{B(a_c-a)/(Aa_0)},
 \nonumber \\
 y_A-y_c=-\sqrt{B(a_c-a)/(Aa_0)}.
 \label{surf39}
 \end{eqnarray}
Thus the equilibrium branch of the curve $y_A(a),y_H(a)$ near the point $a_c,y_c$ is a parabola.

\begin{figure}
    \includegraphics[width=1\linewidth]{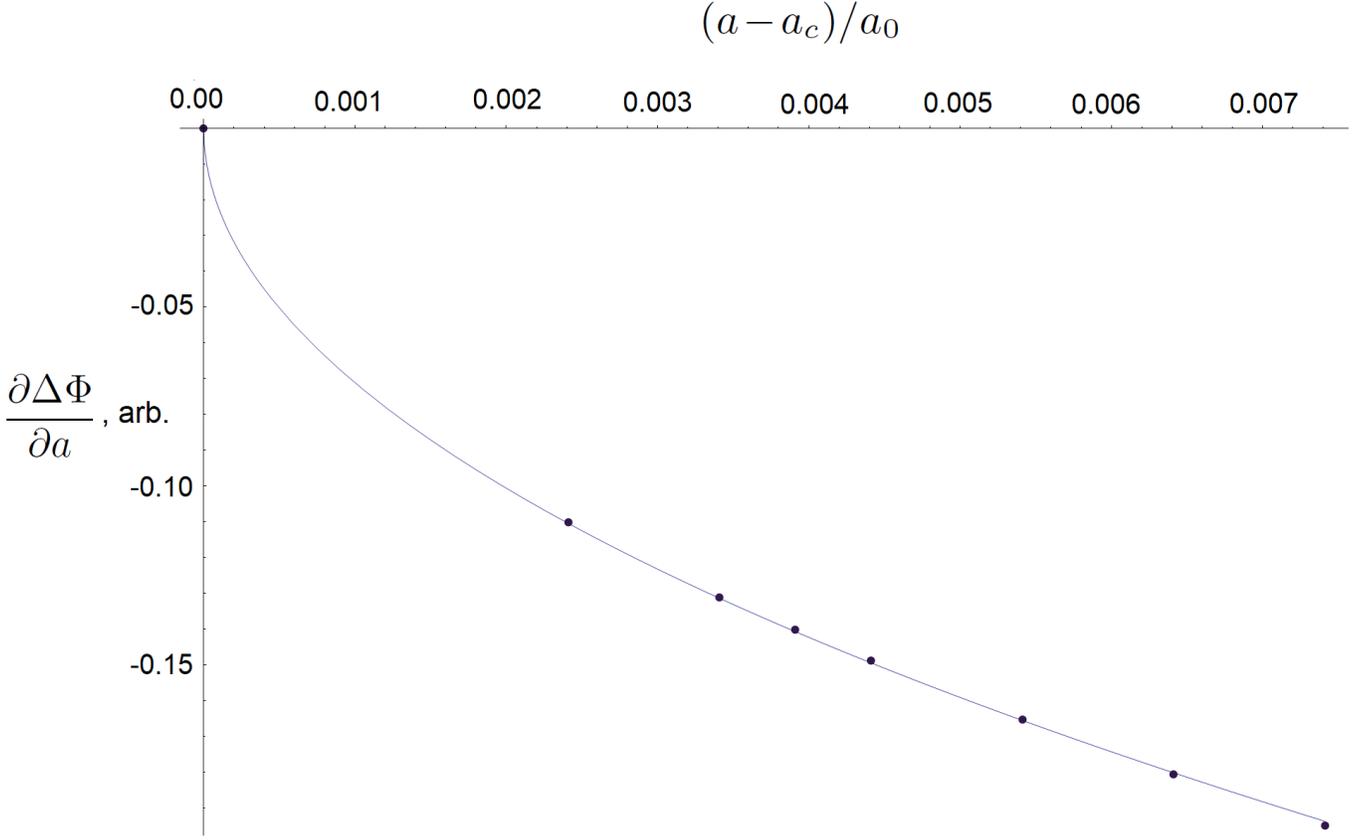}
    \caption{
Comparison of our numeric results (solution to the Eq.(\ref{surf22}) for  ${\partial\Delta \Phi}/{\partial a}\,$
versus $(a-a_c)/a_0$ ) in the coexistence regime (dark blue circles) with analytic theory (Eq. (\ref{surf41}))
near the point $a=a_c$ (blue solid line).
             \label{dPh}
              }
              \end{figure}

Since in the equilibrium the derivatives of $\Delta\Phi$ over $\gamma_A=(\psi_0^2/\xi^2)y_A$ and
$\gamma_H=(\psi_0^2/\xi^2)y_H$ are zero, we find in the main approximation from Eq. (\ref{surf35})
 \begin{equation}
 \frac{\partial\Delta\Phi}{\partial a}
 =2\frac{\psi_0^2}{\xi^2}
 \frac{\partial\Xi}{\partial a}
 \sqrt\frac{B(a_c-a)}{Aa_0},
 \label{surf40}
 \end{equation}
at the equilibrium curve. Here the derivative $\partial\Xi/\partial a$ is taken at $y=y_c$. We conclude that
 \begin{equation}
 \frac{\partial \Delta\Phi}{\partial a}
 \propto (a_c-a)^{1/2},
  \label{surf41}
 \end{equation}
that is the derivative tends to zero as $a\to a_c$, see Fig. \ref{dPh}. Thus, in the agreement with Eqs. (\ref{interval}) the width of the
equilibrium phase coexistence region shrinks, $T_+-T_-\propto (a_c-a)^{1/2}$ as $a\to a_c$, see Figs. \ref{FigureDT}, \ref{Figure2}.
A similar procedure can be used to calculate the higher order terms in $a-a_c$, $y-y_c$ to the expansion (\ref{surf34}).
Technical details and final results are presented in Appendix \ref{sec:thin}, see also Figs. \ref{Ltda}, \ref{Lg}.

\section{Conclusion}
\label{sec:con}

In summary, we developed the theory describing features of the free standing smectic films in the temperature range where the
equilibrium phase coexistence $Sm A$ -- $Hex B$ occurs. Our results explain how the surface induced ordering reduces the width of
the equilibrium phase coexistence region. Quite remarkably the width shrinks to zero, where the film thickness $L$ becomes of the
order of the hexatic correlation length. The behavior of the film at $L \to L_{c}$ resembles the classical gas-liquid critical point, where
the coexisting phases become indistinguishable. Our analysis of the surface-bulk ordering interplay predicts exclusive laws for the equilibrium
phase coexistence range, in terms of the reduced parameters. The described phenomena (and the calculated specific relations
between the parameters) are universal, appropriately rescaled our main predictions depend only on a few dimensionless parameters.
Thus we arrived at the universal picture in terms of the reduced parameters.

Let us stress that our crucial assumption, that the $Sm A$ -- $Hex B$ transition is close to the tricritical point, is strongly supported by
the existing experimental data. For example, see Refs. \cite{JV95,HK97,RD05,MP13,ZK18}, that demonstrates weak first
order phase transitions. Moreover, the measured critical exponents (for the specific heat and for the order parameter) are close to
those for the tricritical point \cite{JV95,HK97,RD05,MP13}. Therefore our theory is applicable to all the materials, and our
predictions (the finite temperature range for the equilibrium phase coexistence, the film thickness as the parameter governing
the width of the coexistence region and universal laws for the width dependence on the system parameters) hold.

We neglected fluctuations of the order parameter. It is well known, that near the tricritical point fluctuations provide
logarithmic corrections to the mean-field values. Since, in accordance with our scheme, in the range of the equilibrium phase coexistence
the control parameter $a$ varies in a relatively narrow interval (on the order of the bulk value $a_0$),
the logarithmic renormalization of the coefficients is not essential for our consideration. However, if
the reducing film thickness approaches the critical value $L_c$, then the smectic and the hexatic states become indistinguishable,
signaling about a special critical point. This special critical point is basically similar to the conventional liquid -- gas critical point,
where fluctuations of the two-component hexatic order parameter (modulus and phase) are relevant (see \cite{PP79} in addition to \cite{LL80,ST87,AN91}). We defer an investigation of the point for a future work.

\begin{figure}
    \includegraphics[width=0.8\linewidth]{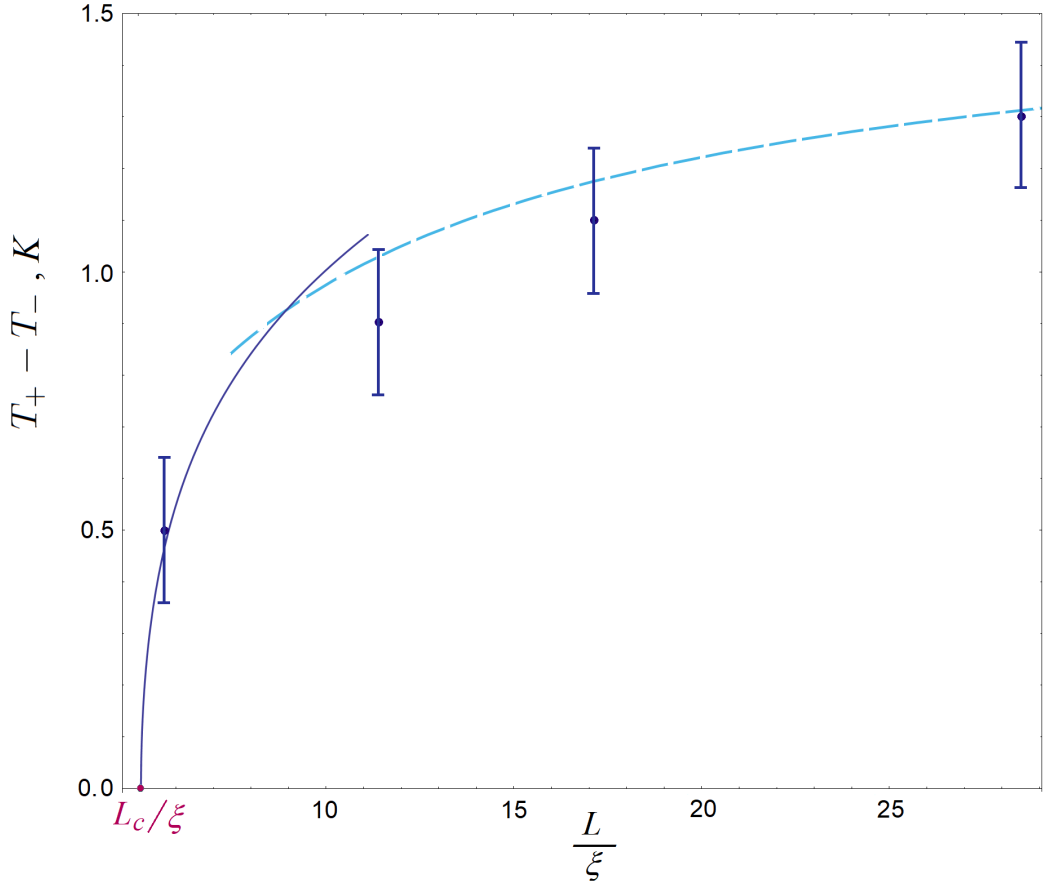}
    \caption{
Borrowed from \cite{ZK18} experimental data on the temperature range $\Delta T$ of the $Sm A$ and $Hex B$ phase coexistence
  as a function of the dimensionless film thickness $L/\xi\,$ (on cooling, shown by dark blue circles).
  Our analytical  description of the data by Eq. (\ref{differ103}) for the  thick films  is shown by the azure dash line in the right.
   In numeric computation we used the following fitting parameters: $\xi =  3.5 \cdot 10^{- 7}$ m, ${\beta \psi_0^2}/\Gamma = 1.6$.
   Our results  near the point $a=a_c$ (the violet solid line) is plotted by solving
    Eqs. (\ref{interval},\ref{surf40},\ref{surf40},\ref{surf41},\ref{thin4}).
        \label{Figure2}
    }
\end{figure}

To illustrate how our theory works we re-analyze the experimental data presented in Ref. \cite{ZK18}
for the $Sm A$ -- $Hex B$ coexistence in the free-standing film of the 54COOBC material.
Measured in \cite{ZK18} the temperature width $\Delta T$ of the phase coexistence region at different
film thickness can be reasonably described by our theory. The comparison suggests also that these experimental data
correspond to the regime of the intermediate film thicknesses (in-between described analytically the thick and thin films
limits). We presented in Fig. \ref{Figure2} (similarly as it has been done in Fig. \ref{Ltda}) our numeric solution to Eqs.
(\ref{surf23},\ref{surf25}), and Eq. \ref{surf8}).

In this work we had deal only with $Sm A$ -- $Hex B$ phase transition in a vicinity of the tricritical point, characterized by the two-component (complex) order parameter. Generally, our theory can be applied to other orientation phase transitions in smectics (provided the
state is close to a tricritical point). For example, it is applicable to the transition between the untilted $Sm A$ and the tilted $Sm C$ states.
However, the explicit expressions require some modifications. Namely, one has to include into consideration, uniaxial
orientational anisotropy within the smectic layers (to compare with the hexagonal symmetry of the $Hex B$ layers), and,
more important, induced by cooperative molecular tilting the layer thickness variation at the transition.

Our theory can be adjusted to describe the paraelectric -- ferroelectric  phase transitions in the solid films as well, where the transition
is close to a tricritical point (see, e.g., \cite{Gerzanich,Strukov} for the case of thin ceramic ferroelectric films). Furthermore for the thin
ferroelectric films surface ordering occurs prior the bulk one, and it yields to a sort critical point, mentioned in \cite{Scott,Duiker,Qu}.
To modify our theory for the ferroelectric solid films, one has to include elastic energy, long-range dipolar forces, domain structures and so on.
Notes also that the equilibrium phase coexistence, tricritical behavior, and the film finite thickness effects are very common in nature, not only
for the smectics or the ferroelectrics, but for spin-density waves, charge density waves, adsorbed atoms as well.

A remarkable peculiarity of Landau theory is that it is a powerful tool for description of different systems in terms of the order
parameter irrespective of its microscopic nature. The system properties depend solely on the system dimension, symmetry, and on the number
of the components of the order parameter. Similarity in the description can be even more close if one considers quasi two-dimensional
layered structures, such as high-temperature superconductors with puzzling properties. What can be useful for us considering other systems?
The matter is that in the smectic liquid crystals, unlike superconductors and superfluids, not only both components of the order parameter have
a transparent physical nature, but also the fields conjugated to the modulus and phase have realistic physical sources (e.g., uniaxial pressure
or electric and magnetic fields). This cannot be said about superconducting gap and superfluid density for which there is no conjugated physical
field. It is tempting to use smectic phases for modeling of different unusual superstructures forming in superconductors and superfluids.
To the same point, the idea (we are advocating here)  on the bulk - surface orderings correspondence in smectic films, became recently very popular with   a number of fascinating applications in several branches of physics, like holographic principle in high energy physics,
or in topological insulators (see, e.g., \cite{BO02,HK10}).

\acknowledgments

This work was inspired by recent X-ray studies of the smectic A and hexatic B phase coexistence in the free standing smectic films \cite{ZK18}.
We are grateful to all members of the experimental team for providing us with the very first results of their remarkable observations.
Special thanks are due to B.I. Ostrovskii, I.A. Vartanyants, I.A. Zaluzhnyy and R.P. Kurta for stimulating discussions. The reported study
was supported by the Ministry of Science and Higher Education of Russia within the State assignment (theme No. 0033-2019-0003).

\appendix

\section{}
\label{sec:thick}

Here we analyze the case of thick films, $L\gg\xi$. Then $a$ is close to $a_0$. The system of equations can be brought
in a more elegant form (ready for numerics) by introducing  dimensionless variables
 \begin{eqnarray}
 \varpi=a/a_0-1, \quad x=\psi^2/\psi_0^2,
 \label{surf29}
 \end{eqnarray}
one obtains
 \begin{eqnarray}
 g=(\psi_0/\xi)^2 x (1+\varpi -2 x+x^2).
 \label{surf4}
 \end{eqnarray}
The parameter $\varpi>0$ is small in our case. The quantity $g$ (\ref{surf4}) has the minimum at $x=x_m$, where
 \begin{equation}
  x_m=\frac{2}{3}\left(
 1+\frac{1}{2}\sqrt{1-3\varpi}\right)<1.
 \label{differ2}
 \end{equation}

As we explained, in the case $L/\xi\gg1$ both, $\psi_A$ and $\psi_H-\psi_m$, are exponentially small in $L/\xi$. Therefore at
analyzing effects, power in $\xi/L$, one can put $\psi_A=0$, $\psi_H=\psi_m =\psi_0\sqrt x_m$. Then one finds from Eq. (\ref{surf5})
 \begin{eqnarray}
 \frac{\xi}{b\psi_0^2}\Phi_A=
 \varphi_A, \qquad
 \label{differ4} \\
 \frac{\xi}{b\psi_0^2}\Phi_H=
 2x_m^2(1-x_m)\frac{L}{\xi} +\varphi_H.
 \label{differ5}
 \end{eqnarray}
The dimensionless quantities $\varphi_A$ and $\varphi_H$ are defined as
 \begin{eqnarray}
 \varphi_A =2\int_0^s
 {dx}\sqrt{x^2-2x+1+\varpi}
 \nonumber \\
 =(s-1)(s^2-2s+1+\varpi)^{1/2}+\sqrt{1+\varpi}
 \nonumber \\
 -\varpi\ln \frac
 {1-s+(s^2-2s+1+\varpi)^{1/2}}{1+\sqrt{1+\varpi}},
 \label{differ1}
 \end{eqnarray}
and
 \begin{eqnarray}
 \varphi_H=2\int_{x_m}^s
 \frac{dx}{\sqrt x}\sqrt
 {x-2(1-x_m)}\, (x-x_m)=
 \nonumber \\
 (s-x_m-1)\sqrt{s(s-2+2x_m)}
 +\sqrt{x_m(3x_m-2)}
 \nonumber \\
 +2\varpi \ln \frac
 {\sqrt s+\sqrt{s-2(1-x_m)}}
 {\sqrt {x_m}+\sqrt{3x_m-2}},
 \label{differ3}
 \end{eqnarray}
where the subscript $s$ corresponds to the surface value of the order parameter.

Using Eqs. (\ref{differ1},\ref{differ3}), one can easily calculate
 \begin{eqnarray}
 \lim_{s\to\infty}(\varphi_A-\varphi_H)=
 \sqrt{1+\varpi} -\sqrt{x_m(3x_m-2)}
 \nonumber \\
 +x_m-\varpi\ln\frac
 {2(\sqrt{1+\varpi}\,-1)}{(\sqrt {x_m}+\sqrt{3x_m-2})^2}.
 \label{differ}
 \end{eqnarray}
Therefore the condition $\Delta\Phi=0$ reads as
 \begin{eqnarray}
 \sqrt{1+\varpi}
 -\sqrt{x_m(3x_m-2)} +x_m
 \nonumber \\
 -\varpi\ln\frac
 {2(\sqrt{1+\varpi}\,-1)}{(\sqrt {x_m}+\sqrt{3x_m-2})^2}
 \nonumber \\
 =2x_m^2(1-x_m)\frac{L}{\xi}.
 \label{differ6}
 \end{eqnarray}
This equation relates $\xi/L$ and $\varpi$.

Now we turn to the relation (\ref{surf19}) that can be rewritten as
 \begin{eqnarray}
 \frac{\Gamma}{\beta \psi_0^2}(T_+-T_-)
 +\frac{\xi}{L}\int_0^s \frac{dx}{\sqrt{x^2-2x+1+\varpi}}
 \nonumber \\
 -\frac{\xi}{L}\int_{x_m}^s \frac{dx}{\sqrt{x^2-2(1-x_m)x}}
 -x_m=0.
 \label{differ7}
 \end{eqnarray}
The integrals here are
 \begin{eqnarray}
 \int_0^s \frac{dx}{\sqrt{x^2-2x+1+\varpi}}
 =\ln\frac{s-1+\sqrt{s^2-2s+\varpi}}{\sqrt{1+\varpi}\,-1},
 \nonumber \\
 \int_{x_m}^s \frac{dx}{\sqrt{x^2-2(1-x_m)x}}
 =2\ln\frac{\sqrt s+\sqrt{s-2+2x_m}}{\sqrt {x_m}+\sqrt{3x_m-2}}.
 \nonumber
 \end{eqnarray}
Substituting the expressions into Eq. (\ref{differ7}) and passing to the limit $s\to\infty$, one obtains
 \begin{eqnarray}
  \frac{\Gamma(T_+-T_-)}{\beta \psi_0^2}
 +\frac{\xi}{L}
 \ln\frac{(\sqrt {x_m}+\sqrt{3x_m-2})^2}{2(\sqrt{1+\varpi}\,-1)}
 =x_m. \quad
 \label{differ8}
 \end{eqnarray}
The equation relates $\xi/L$ and $T_+-T_-$.

The expressions (\ref{differ2},\ref{differ6},\ref{differ8}) admit a regular expansion in $\varpi$. Keeping zero and first terms of
the expansion, we get
 \begin{eqnarray}
 \frac{1}{\varpi}+\ln\frac{4}{\varpi}
 +\frac{5}{4}=\frac{L}{\xi},
 \label{differ9} \\
 \frac{\Gamma(T_+-T_-)}{\beta \psi_0^2}=
 \frac{\xi}{L\varpi}+\frac{\xi}{L}-\frac{\varpi}{4}.
 \label{differ10}
 \end{eqnarray}
Taking into account only the main logarithmic term, we reproduce Eqs. (\ref{surf18},\ref{surf20}). In the next order in $\varpi$
one finds the relations
 \begin{eqnarray}
 \frac{1}{\varpi} + (1 + \frac{\varpi}{4})\, \ln\frac{4}{\varpi}
 +\frac{5}{4} + \frac{\varpi}{16}=\frac{L}{\xi},
 \label{differ91} \\
 \frac{\Gamma(T_+-T_-)}{\beta \psi_0^2}=
 \frac{\xi}{L\varpi}+\frac{\xi}{L}+\frac{3\varpi}{8}\frac{\xi}{L}
 -\frac{\varpi}{4}- \frac{\varpi^2}{4}.
 \label{differ101}
 \end{eqnarray}

Expressing $\varpi$ via $\xi/L$ from Eq. (\ref{differ91}), we obtain in the same approximation
 \begin{eqnarray}
 \varpi= \frac{\xi}{L}+
 \left(\frac{\xi}{L}\right)^2
 \left(\ln\frac{4L}{\xi}+\frac{5}{4}\right)
 \nonumber \\
 +\left(\frac{\xi}{L}\right)^3
 \left[\left(\ln\frac{4L}{\xi}\right)^2
 +\frac{7}{4}\ln\frac{4L}{\xi}+\frac{3}{8}\right],
 \label{differ102}
 \end{eqnarray}
the function ${L}/{\xi}$ versus $\varpi$ is presented in Fig. \ref{Ltda}. Substituting the expression (\ref{differ102}) into
Eq. (\ref{differ101}), we finally find
 \begin{eqnarray}
 \frac{\Gamma(T_+-T_-)}{\beta \psi_0^2}=
 1-\frac{\xi}{L}\left(\ln\frac{4L}{\xi}+\frac{1}{2}\right)
 \nonumber \\
 +\left(\frac{\xi}{L}\right)^2
 \left(\frac{1}{2}\ln\frac{4L}{\xi}+1\right),
 \label{differ103}
 \end{eqnarray}
in the second order in $\xi/L$. We plot the corresponding dependence of $T_+-T_-$  on the dimensionless film thickness $L/\xi\,$
in Fig. \ref{FigureDT}.

\section{}
\label{sec:thin}

Here we analyze in more detail the case where $L$ is close to $L_c$ and the coexistence region is rather narrow in its width.
Then one should start from the expression (\ref{surf34}), correct near the point $a_c,y_c$. We discuss next corrections to the
expression (\ref{surf34}). The modified expression can be written as
 \begin{eqnarray}
 \frac{\Xi}{\xi}=\frac{\Xi(a,y_c)}{\xi}
 -A(y-y_c)^3
 -B \frac{a-a_c}{a_0}(y-y_c)
 \nonumber \\
 +C(y-y_c)^4
 +D\frac{a-a_c}{a_0}(y-y_c)^2, \qquad
 \label{thin1} \\
 \frac{\Xi(a,y_c)}{\xi}
 =\frac{\Xi_c}{\xi}
 +A_1\frac{a-a_c}{a_0}
 +B_1\left(\frac{a-a_c}{a_0}\right)^2, \quad
 \label{thin0}
 \end{eqnarray}
where $A,B,D,C,A_1,B_1$ are dimensionless parameters. The corrections with the coefficients $C,D$ contain an
extra power of $y-y_c$ in comparison with the main terms with the coefficients $A,B$. The parameters $D,C,A_1,B_1$ can be
found numerically, they are $D=-45.6325$, $C=-724.459$, $A_1=-4.81157$, $B_1=14.4096$.

\begin{figure}
    \includegraphics[width=1\linewidth]{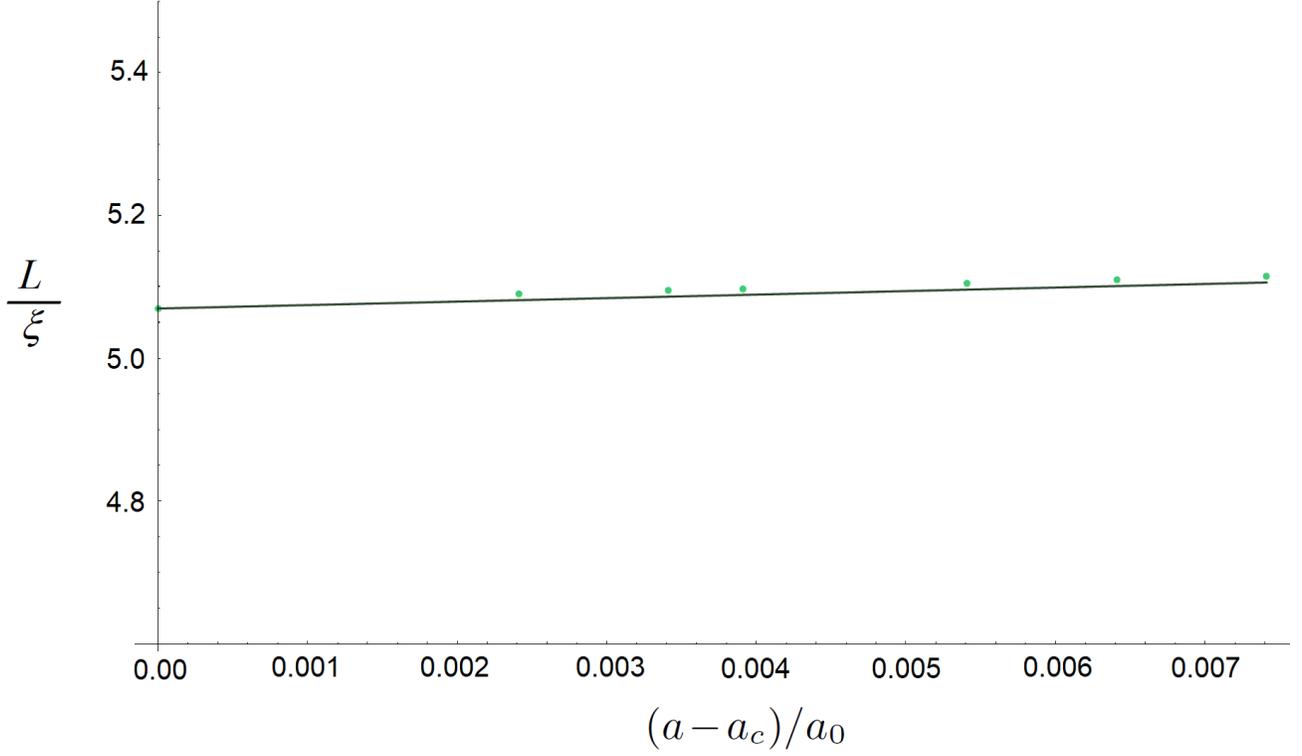}
    \caption{
Comparison of the numeric results for $L/\xi\,$ (light green circles) versus $a-a_c$ with their analytic counterparts given by
Eq.((\ref{thin4})) near the point $a=a_c$. Branch corresponding to the  Eq.(\ref{thin4}) is shown by the dark green solid line.
             \label{Lda}
              }
              \end{figure}
\begin{figure}
    \includegraphics[width=0.9\linewidth]{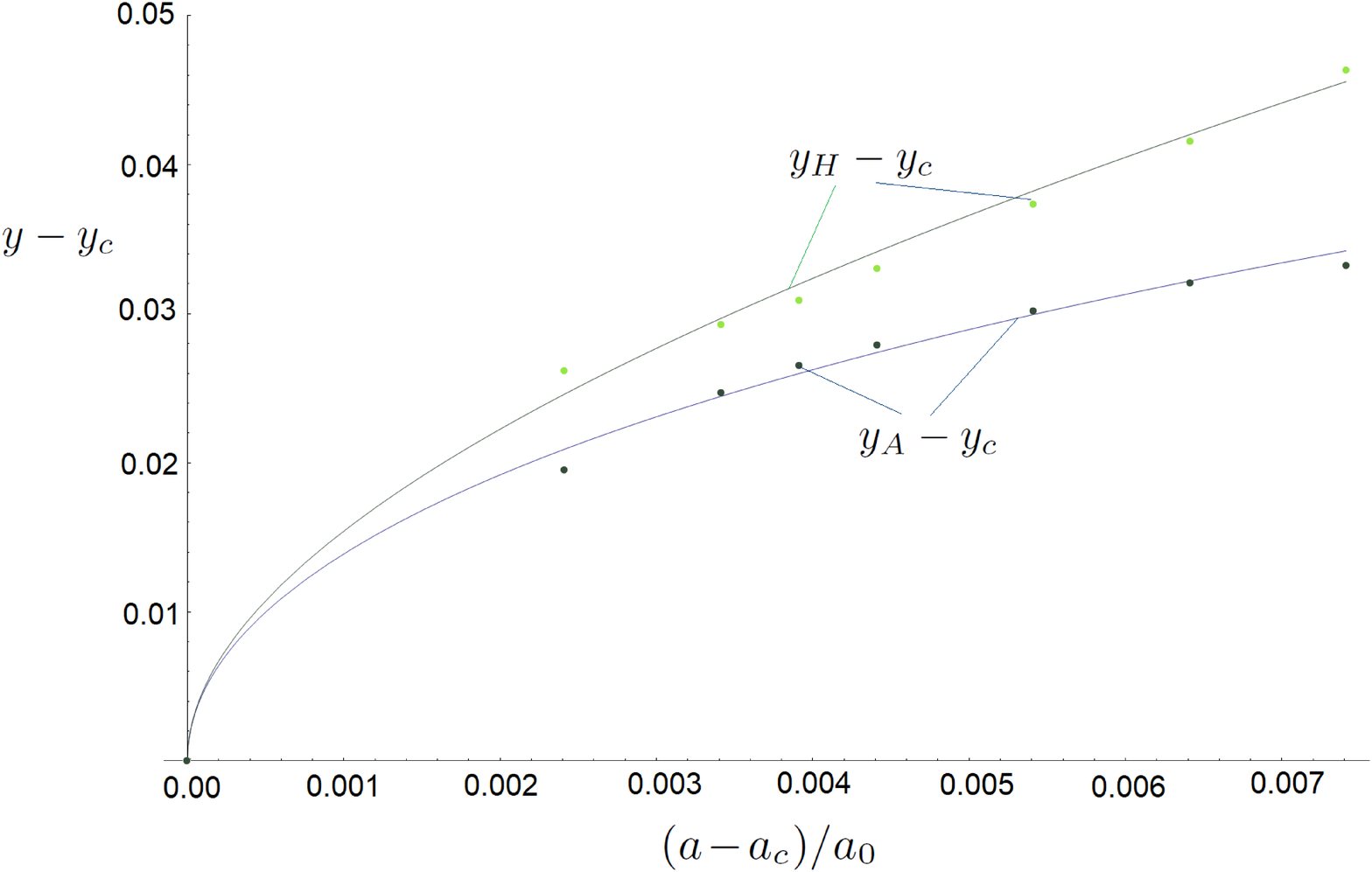}
    \caption{
Comparison of the numeric results for  $\,y_H-y_c\,$ (light green circles) and $\,y_A-y_c\,$
(blue circles) versus $a-a_c$ with analytic  ones given by  Eqs.((\ref{thin5})) near the point $a=a_c$.
Curves corresponding to the Eqs. (\ref{thin5}) are shown by the solid lines (upper dark green for $Hex B$ phase and bottom blue for $Sm A$) .
                   \label{gphi}
              }
              \end{figure}

The next step is in generalizing Eq. (\ref{surf35})
 \begin{eqnarray}
 \frac{\xi\Delta\Phi}{\psi_0^2}
 =\left[\frac{\Xi(a,y_c)}{\xi}-\frac{L}{\xi}\right](y_H-y_A)
 \nonumber \\
 -\frac{A}{4}\left[
 (y_H-y_c)^4-(y_A-y_c)^4 \right]
 \nonumber \\
 -\frac{B}{2}
 \frac{a-a_c}{a_0}
 \left[(y_H-y_c)^2-(y_A-y_c)^2 \right]
 \nonumber \\
 +\frac{C}{5}\left[
 (y_H-y_c)^5-(y_A-y_c)^5 \right]
 \nonumber \\
 +\frac{D}{3}(\varpi-\varpi_c)\left[
 (y_H-y_c)^3-(y_A-y_c)^3 \right].
 \label{thin2}
 \end{eqnarray}
Now we can find the equilibrium values of the parameters that are determined by the conditions
(\ref{surf8}) and $\Delta\Phi=0$. The conditions (\ref{surf8}) are written as
 \begin{eqnarray}
 \Xi(a,y_c)/\xi-L/\xi
 \nonumber \\
 =A(y_H-y_c)^3+B(\varpi-\varpi_c) (y_H-y_c)
 \nonumber \\
 +C(y_H-y_c)^4
 +D\frac{a-a_c}{a_0}
 (y_H-y_c)^2
 \nonumber \\
 =A(y_A-y_c)^3+B\frac{a-a_c}{a_0}(y_A-y_c)
 \nonumber \\
 +C(y_A-y_c)^4
 +D\frac{a-a_c}{a_0}(y-y_c)^2.
 \label{thin3}
 \end{eqnarray}
The expressions generalize Eq. (\ref{surf37}). The condition $\Delta\Phi=0$ gives the equation following from Eq. (\ref{thin2}).

To have a regular expansion (perturbation theory) we assume the higher order corrections to be small. Then we find
 \begin{eqnarray}
 \frac{L}{\xi}-\frac{\Xi(a,y_c)}{\xi}
 =\left(\frac{C}{5}-\frac{DA}{3B}\right)
 \left(\frac{B(a-a_c)}{Aa_0}\right)^2,
 \label{thin4} \\
 y_H,y_A=y_c\pm \sqrt{\frac{B(a-a_c)}{Aa_0}}
 \nonumber \\
 -\left(\frac{3C}{5A}-\frac{2D}{3B}\right)
 \frac{B}{A}\frac{a-a_c}{a_0},
 \label{thin5}
 \end{eqnarray}
instead of Eqs. (\ref{surf38},\ref{surf39}). The applicability condition of the expressions implies that the corrections to $y_H,y_A$
are small in comparison with the main contribution. Comparing the expression (\ref{thin4}) with Eq. (\ref{thin0}),
we conclude, that $L$ is expanded over integer powers of $(a-a_c)/{a_0}$. Our numeric results, shown in Figs. \ref{Lda}, \ref{gphi}
are in agreement with presented above analytic expansion, see Eqs. (\ref{thin4},\ref{thin5}).

\end{document}